\documentclass[conference]{IEEEtran}
\IEEEoverridecommandlockouts

\usepackage{times}

\usepackage{dsfont}
\usepackage{cite}
\usepackage[font=small]{caption}
\usepackage[font=small]{subcaption}
\usepackage{mathtools}
\usepackage{multirow}
\usepackage{multicol,lipsum}
\usepackage{mathdots}
\usepackage{yhmath}
\usepackage{cancel}
\usepackage{color}
\usepackage{array}
\usepackage{tabularx}
\usepackage{booktabs}
\usepackage{tikz}
\usetikzlibrary{automata, positioning, arrows,shapes.multipart,fit}
\usetikzlibrary{shapes}
\usetikzlibrary{fadings}
\usetikzlibrary{patterns}
\usetikzlibrary{shadows.blur}
\usepackage{latexsym}
\usepackage{graphicx}
\usepackage{verbatim}
\usepackage{url}
\usepackage{pst-node}
\usepackage{latexsym}
\usepackage{amsmath,amsfonts,amssymb}
\usepackage{bm}
\usepackage{algorithmic}
\usepackage[linesnumbered,ruled,vlined]{algorithm2e}

\SetCommentSty{mycommfont}
\usepackage{graphicx,epstopdf}

\makeatletter
\newcommand*{\rom}[1]{\expandafter\@slowromancap\romannumeral #1@}
\makeatother

\usepackage{epsfig}
\usepackage{siunitx}

\usepackage[nolist,nohyperlinks]{acronym}

\usepackage{hyperref}
\usepackage{epstopdf}
\usepackage{tabulary}
\usepackage{multirow}
\usepackage{makecell}
\usepackage{stfloats}  
\usepackage{longtable}

\def\bb0{{\mathbb{0}}}

\def\bb{{\mathbf{b}}}

\def\bv{{\mathbf{v}}}

\def\b0{{\mathbf{0}}}

\def\bH{{\mathbf{H}}}

\def\bX{{\mathbf{X}}}

\def\bOmega{{\mathbf{\Omega}}}

\def\bbR{{\mathbb{R}}}

\def\cB{\mathcal{B}}
\def\cC{\mathcal{C}}

\def\cJ{\mathcal{J}}
\def\cK{\mathcal{K}}

\def\cS{\mathcal{S}}

\def\sfT{\mathsf{T}}

\def\sf0{{\mathsf{0}}}

\def\1m{\textrm{1 m}}

\newcommand{\snr}{{\mathrm{SNR}}}

\DeclareMathOperator{\argmax}{arg\,max} 

\DeclarePairedDelimiter\abs{\lvert}{\rvert} 
\newcommand\norm[1]{\left\lVert#1\right\rVert}
\newcommand{\var}{\texttt}
\DeclareMathAlphabet\mathbfcal{OMS}{cmsy}{b}{n}

\begin{document}
	
	\title{Performance of Predictive Indoor mmWave Networks with Dynamic Blockers}
    \author{
    {Andrea~Bonfante, Lorenzo~Galati~Giordano, Irene Macaluso and~Nicola~Marchetti}%
    \thanks{A. Bonfante, I. Macaluso and  N. Marchetti are with CONNECT Centre, Trinity College Dublin, Ireland (e-mail: bonfanta@tcd.ie; nicola.marchetti@tcd.ie; MACALUSI@tcd.ie).
    L. Galati Giordano is with Nokia Bell Labs, Stuttgart, Germany (e-mails: lorenzo.galati\_giordano@nokia-bell-labs.com). 
    This work was supported by Irish Research Council and by Nokia Ireland Ltd under Grant Number EPSPG/2016/106.}}%
	\maketitle
	\markboth{DRAFT}{Shell \MakeLowercase{\textit{et al.}}: Bare Demo of IEEEtran.cls for IEEE Journals}
	\begin{abstract}
	In this paper, we consider millimeter Wave (mmWave) technology to provide reliable wireless network service within factories where links may experience rapid and temporary fluctuations of the received signal power due to dynamic blockers, such as humans and robots, moving in the environment. 
	We propose a novel beam recovery procedure that leverages Machine Learning (ML) tools to predict the starting and finishing of blockage events. This erases the delay introduced by current 5G New Radio (5G-NR) procedures when switching to an alternative serving base station and beam, and then re-establish the primary connection after the blocker has moved away.
	Firstly, we generate synthetic data using a detailed system-level simulator that integrates the most recent 3GPP 3D Indoor channel models and the geometric blockage Model-B. Then, we use the generated data to train offline a set of beam-specific Deep Neural Network (DNN) models that provide predictions about the beams' blockage states. Finally, we deploy the DNN models online into the system-level simulator to evaluate the benefits of the proposed solution. Our prediction-based beam recovery procedure guarantee higher signal level stability and up to $82\%$ data rate improvement with respect detection-based methods when blockers move at speed of $2$ m/s.
	\end{abstract}
	\begin{acronym}[mmWave-BSs] 
\acro{10GE}{10 Gigabit Ethernet}
\acro{3D}{Three-dimensional}
\acro{3GPP}{Third Generation Partnership Project}
\acro{5G}{5-th Generation}
\acro{AC}{Access}
\acro{ACK}{Acknowledgment }
\acro{Adam}{ADAptive Moment estimation}
\acro{AoA}{Angle of Arrival}
\acro{AP}{Access Point}
\acro{AoD}{Angle of Departure}
\acro{AR}{Augmented Reality}
\acro{BER}{Bit Error Rate}
\acro{BF}{Beamforming}
\acro{BH}{Backhaul}
\acro{BLAS}{Basic Linear Algebra Subprograms}
\acro{BS}{Base Station}
\acro{BSs}{Base Stations}
\acro{CDF}{cumulative distribution function}
\acro{CP}{Cyclic Prefix}
\acro{CPU}{Central Processing Unit }
\acro{CSI}{Channel State Information}
\acro{DL}{Downlink}
\acro{DNN}{Deep Neural Network}
\acro{DoA}{Direction of Arrival}
\acro{DoD}{Direction of Departure}
\acro{DSP}{Digital Signal Processor}
\acro{E2E}{End-to-end}
\acro{eMBB}{Enhanced Mobile BroadBand}
\acro{FFT}{Fast Fourier Transform}
\acro{FN}{False Negative}
\acro{FP}{False Positive}
\acro{FPGA}{Field-Programmable Gate Array}
\acro{GT}{Ground Truth}
\acro{HD}{Half-Duplex}
\acro{HetNet}{Heterogeneos Network}
\acro{HW}{Hardware}
\acro{IID}{Independent and Identically Distributed}
\acro{ISD}{Inter Site Distance}
\acro{L3}{Layer 3}
\acro{LoS}{Line-of-Sight}
\acro{LP}{Linear Program}
\acro{LSAS}{Large Scale Antenna System}
\acro{LTE}{Long Term Evolution}
\acro{MAC}{Medium Access Control}
\acro{MCS}{Modulation and Coding Scheme}
\acro{MILP}{Mixed Integer Linear Program}
\acro{MIMO}{Multiple-Input-Multiple-Output}
\acro{MKL}{Math Kernel Library}
\acro{ML}{Machine Learning}
\acro{MLP}{Multilayer Perceptron}
\acro{mmWave}{millimeter Wave}
\acro{mmWave-BS}{mmWave Base Station}
\acro{mmWave-BSs}{mmWave Base Stations}
\acro{MNO}{Mobile Network Operator}
\acro{MPC}{Multipath Components}
\acro{MWC}{Mobile World Congress}
\acro{NLoS}{Non Line-of-Sight}
\acro{NR}{New Radio}
\acro{OFDM}{Orthogonal Frequency Division Multiplexing}
\acro{PHY}{Physical layer}
\acro{RAN}{Radio Access Network}
\acro{RB}{Resource Block}
\acro{ReLu}{Rectified Linear unit}
\acro{RF}{Radio Frequency}
\acro{RLF}{Radio Link Failure}
\acro{RSRP}{Reference Signal Received Power}
\acro{RSRP}{Reference Signal Received Power}
\acro{RSSI}{Received Signal Strength Indicator}
\acro{Rx}{Receiver}
\acro{SA}{standalone}
\acro{s-BH}{self-Backahuling}
\acro{SC}{small-cell}
\acro{SCM}{Spatial Channel Model}
\acro{SGD}{Stochastic Gradient Descent}
\acro{SNR}{signal-to-noise ratio}
\acro{SSB}{synchronisation signal block}
\acro{SVN}{Subversion}
\acro{SW}{Software}
\acro{TDD}{Time Division Duplex}
\acro{TI}{Texas Instruments}
\acro{TN}{True Negative}
\acro{TP}{True Positive}
\acro{TTI}{Transmission Time Interval}
\acro{Tx}{Transmitter}
\acro{UDP}{User Datagram Protocol}
\acro{UE}{User Equipment}
\acro{UL}{Uplink}
\acro{UPA}{Uniform Planar Array}
\acro{URLLC}{Ultra-Reliable Low-Latency Communication}
\acro{VR}{Virtual Reality}
\acro{WLAN}{Wireless Local Area Network}
\end{acronym}
	\section{Introduction}
	\IEEEpubidadjcol
	
	\IEEEPARstart{T}{he} integration of \ac{ML} tools with wireless communication systems is envisioned as one of the essential steps towards future intelligent \acp{RAN} \cite{8542764,6489878}. 
    \ac{ML} tools can be employed to design novel resource management and control methods to optimise the performance of \ac{5G} and beyond wireless systems \cite{8743390}. 
    One of the key technologies for \ac{5G} networks is \ac{mmWave} communication, where the large available bandwidth at \ac{mmWave} frequencies can support extremely high capacity wireless links, offering data rates in the order of gigabits per second \cite{6732923,6824746,6932503}. 
    MmWave communications are well-suited for indoor network deployments \cite{9094692}, e.g. open offices, shopping centres and industrial spaces. 
    Adopting multi-antenna arrays with analog-only or hybrid architectures enables forming narrow beams that provide high directivity gains and compensate for the severe pathloss \cite{7342886}. 
    Nevertheless, in this scenario, the \ac{mmWave} links experience rapid and temporary fluctuations of the received signal power when they encounter blocking objects, such as humans and robots moving in the environment \cite{8254900}. 
    The links' data rate may become highly intermittent, making it highly challenging to guarantee high data rate and low latency services \cite{8446037,7876982}. 

    In general, the blockage effect follows non-periodic time dynamics and cannot be predicted in advance from the received power using time series forecasting models, as there is no indication that the blockage will happen until the blockage causes a significant change in the received signal power. 
    Therefore, conventional beam recovery methods are based on blockage detection. 
    For instance, the beam failure recovery (BFR) procedure adopted by \ac{3GPP} \ac{NR} standard \cite{3gpp.38.321} and other methods shown in \cite{6134444,8363902} rely on a signal threshold, detecting the blockage events when the received signal power drops below a given value. 
    The beam recovery operations yield a delay as they establish the backup beam after detecting the blockage, leading to a data rate loss every time the blocker intersects the link. 

    On the other hand, context data collected, for example, by using other \ac{mmWave} links \cite{9148975,8824987}, utilising the Sub-6 GHz spectrum bands \cite{9121328,8734054} or by adopting external sensors (e.g. camera \cite{8792137,8941039} and radar \cite{10.1145/2980115.2980134}), 
    provide relevant information about the environment surrounding the \ac{mmWave} link's path. 
    Leveraging these data enables to predict if a blockage event will occur at any given time on the \ac{mmWave} link. 
    Thus, the blockage predictions can trigger the beam recovery operations ahead of the blockage events to avoid the data rate loss experienced by the detection methods. 
    The early works \cite{8792137,8941039,9121328,9148975,8824987} showed promising results when applying \ac{ML} tools to the blockage prediction problem. 
    Nevertheless, their analysis is limited to the single-\ac{UE} scenario, not accounting for multiple \acp{UE}, which would be a more realistic scenario for modelling \ac{mmWave} network deployments and enable exploring the feasibility of applying \ac{ML} predictions at the network level. 

	\subsection{Related Works}\label{Sec:1b}
	
	In this section, we review state-of-the-art approaches that use \ac{ML} tools to make the blockage prediction of \ac{mmWave} links. 
	There are three major techniques regarding the data acquisition for training the \ac{ML} models: (i) using data acquired from the \ac{mmWave} channel measurements, (ii) utilising complementary data from the Sub-6\,GHz channel bands and (iii) collecting data through a camera or radar mounted on top of the \ac{mmWave-BS}. 
	
	The authors of \cite{9148975} consider using multi-link \ac{mmWave} channel measurements as input data for a Long Short-Term Memory (LSTM) model that predicts the received signal power variations of one of the link in advance. 
	In the single-cell scenario, \ac{SNR} measurements related to both \ac{LoS} and \ac{NLoS} mmWave links with the serving \ac{BS} need to be collected, requiring sweeping the beam towards the \ac{NLoS} paths directions and interrupting the ongoing data communication with the \ac{UE}. 
    Moreover, in the multi-cell scenario, the method entails acquiring the \ac{SNR} measurements from the \ac{LoS} paths with neighbour \acp{BS}. 
    In this case, the neighbour \acp{BS} perform beam sweeping, changing the beams towards \acp{UE} that are not served, and provide the additional \ac{SNR} measurements.  
    Thus, integrating the LSTM-based method proposed in \cite{9148975} into the \ac{5G} \ac{NR} \ac{mmWave} network operations can be challenging to realise in practice and is still an open research problem. 
    Differently in \cite{8824987}, the authors propose to use unsupervised online learning to predict the \ac{mmWave} link state. 
    This approach utilises the \ac{MPC} of the \ac{mmWave} channel as input data for the \ac{ML} model; however, the predictions are limited to one blocker and rely on its location information, which is usually not available unless, for instance, the blocker is equipped with a localisation device. 
    Another interesting work that relies on data acquired from the \ac{mmWave} channel measurements is \cite{8646438}, where a \ac{UE} moving in a vehicle utilises the \ac{ML} model predictions to select the \ac{BS} that takes over the communication to avoid the blockage. However, this work is suited for vehicular networks, and we consider this scenario outside the scope of this paper. 
    
    Another line of work focuses on using the Sub-6\,GHz spectrum bands measurements to observe the rapid variations of the \ac{mmWave} channel due to the blockage, tens of ms ahead of the blockage event \cite{8895815}. 
    Following this approach, in \cite{9121328}, the authors propose a \ac{DNN} model that uses the Sub-6\,GHz channel measurements to predict the beam state and the optimal beam of the codebook that maximises the achievable rate. 
    Their results show a $90\%$ probability of correct predictions of the \ac{mmWave} link blockage state. 
    This method is verified with the single-\ac{UE} scenario, and it is impractical for the \ac{SA} mmWave network deployment. 
	
	Alternatively, in \cite{8792137}, the authors propose a \ac{ML}-based predictor of received signal power that uses visual data acquired by a camera pointing to the \ac{BS}-\ac{UE} path. Similarly, in \cite{8941039}, the authors use visual data to predict the \ac{BS} to which handover the connection. 
    Conversely, the RadMAC project is a proof-of-concept that deploys a mmWave radar on top of the mmWave \ac{BS} \cite{10.1145/2980115.2980134}. 
    RadMAC tracks with the radar the presence of moving objects, e.g., humans, that can eventually intersect the \ac{mmWave} link, switching the beam preemptively to a backup link during the blockage events. 
    The applicability of the visual and radar-based systems is proved for the single-\ac{UE} case, where it is possible to cover the beam path serving the \ac{UE} with a wide camera angle or with the radar system. However, practical \ac{mmWave} systems use beam steering with codebook-based transmission to cover multiple \acp{UE} in the sector, making it more challenging to employ visual and radar-based methods to determine the blockage state for each beam. 
	
	\subsection{Contributions}
	
	In this paper, we propose a novel beam recovery method that is integrated with \ac{DNN} models to predict blockage events. 
    This method relies on the early indication of the predictions to switch to an alternative beam pair, slightly before the blockage starts to impact the received signal power. 
    Unlike the works \cite{8792137,8941039,9121328,9148975,8824987}, 
    we consider extending the blockage prediction method to a more realistic multi-\ac{UE} setting, enabling us to deploy the \ac{ML} models into the network to study the data rate performance. 
    Moreover, most of the existing works necessitate extra \ac{HW} components, e.g. camera and Sub-6 GHz transceiver, while we rely exclusively on the \ac{mmWave} channel measurements aggregated from multiple \acp{BS}. 
    The \ac{ML} model proposed in \cite{9148975} uses input data that necessitate additional \ac{SNR} measurements from the ones already defined in the \ac{5G} \ac{NR} specifications.
    Instead, we use standard beam-quality measurements reported by the \acp{UE} to the \acp{BS} according to the \ac{3GPP} \ac{NR} specification \cite{3gpp.38.300}. 
    
    We consider a \ac{mmWave} Indoor network with high \acp{UE} density and multiple \acp{BS}. 
    We account for \acp{BS} employing multi-antenna arrays with analog beamforming architecture and codebook-based signal transmission. 
    We incorporate into the \ac{3GPP}-based system-level simulator a geometric blockage model (Blockage model B) that ensures consistent results over time, space and frequency components of the channel \cite{3gpp.38.901}. 
    The beams intersecting the blocker in locations that are spatially close have beam-quality measurements that follow similar temporal dynamics. 
    Hence, the blocker presence at time $t$ on one beam can indicate the blockage presence on other beams in successive instants of time. 
    Thus, we train a \ac{DNN} model to predict the beam-specific blockage state, taking as input data the other beams' measurements exchanged with the neighbour \acp{BS} through a central controller. 
    Our contributions can be summarised as follows:
    \begin{enumerate}
        \item We train a set of beam-specific \ac{DNN} models with synthetic data generated through a \ac{3GPP}-based system-level simulator. The models exclusively use existing beam-quality measurements reported by the \acp{UE} to the \acp{BS} as indicated in the \ac{5G} \ac{NR} specifications.
    
        \item We propose a novel multi-\ac{UE} prediction-based method for beam recovery that uses the \ac{DNN} models' output to control and initiate the beam switching in advance and complete this operation by the time the \ac{mmWave} link becomes blocked. 
        
        \item We validate the prediction-based method by deploying the \ac{DNN} models online into a \ac{3GPP}-based system-level simulator to verify the close match to the performance of an ideal method that has perfect knowledge of the future beam states. We then compare the prediction method to other two cases: \emph{i)} adopting the beam recovery method based on detection and \emph{ii)} utilising a fixed beam method. 
        We show that the prediction method switches to a backup beam earlier than the blockage event, and avoids the data rate loss that occurs with the detection method due to the delay in switching to the backup beam and without the switching for the fixed beam method. 
        
        \item We provide quantitative data rate results of the three methods varying the blocker speed. 
        During blocked time instants, the prediction-based method improves the 25-th percentile of the fixed-beam and detection-based methods data rates by $238\%$ and $24\%$ for blocker speed of $1$ m/s, and by $223\%$ and $82\%$ for blocker speed of $2$ m/s. This shows the benefits of using the prediction-based method for the beam recovery operation, especially for the worst served \acp{UE} and higher blocker speed. 
    \end{enumerate}

	The remainder of this paper is organised as follows.
	Section \ref{Sec:2} describes the system model and the \ac{mmWave} \ac{DL} transmission;
	Section \ref{Sec:3} introduces the beam recovery method based on detection and the beam recovery method based on predictions; 
	Section \ref{Sec:4} describes the procedure to obtain the beam state predictions; 
	Section \ref{Sec:5} presents the evaluation of the methods under analysis and the main results, 
	and \ref{Sec:6} summarises the key findings and directions for future research. 
 	The following notation is used throughout the paper: boldface lower case and boldface upper case are used for column vectors $\textbf{x}$ and matrices $\textbf{X}$, respectively. 
	$\textbf{x}^{\sfT}$ denote the transpose of $\textbf{x}$. 
	A complex Gaussian random variable $x$ is denoted $x \sim \mathcal{CN}(\mu,\,\sigma^{2})$, where $\mu$ is the mean and $\sigma^{2}$ is the variance. 
	$\left[X \right]$ denotes the Iverson bracket that is defined to be 1 when $X$ is true and 0 when $X$ is false. 
    
    \section{System Model}\label{Sec:2}
    \begin{figure}[t]
		\centering
		\includegraphics[width=0.6\columnwidth,keepaspectratio]{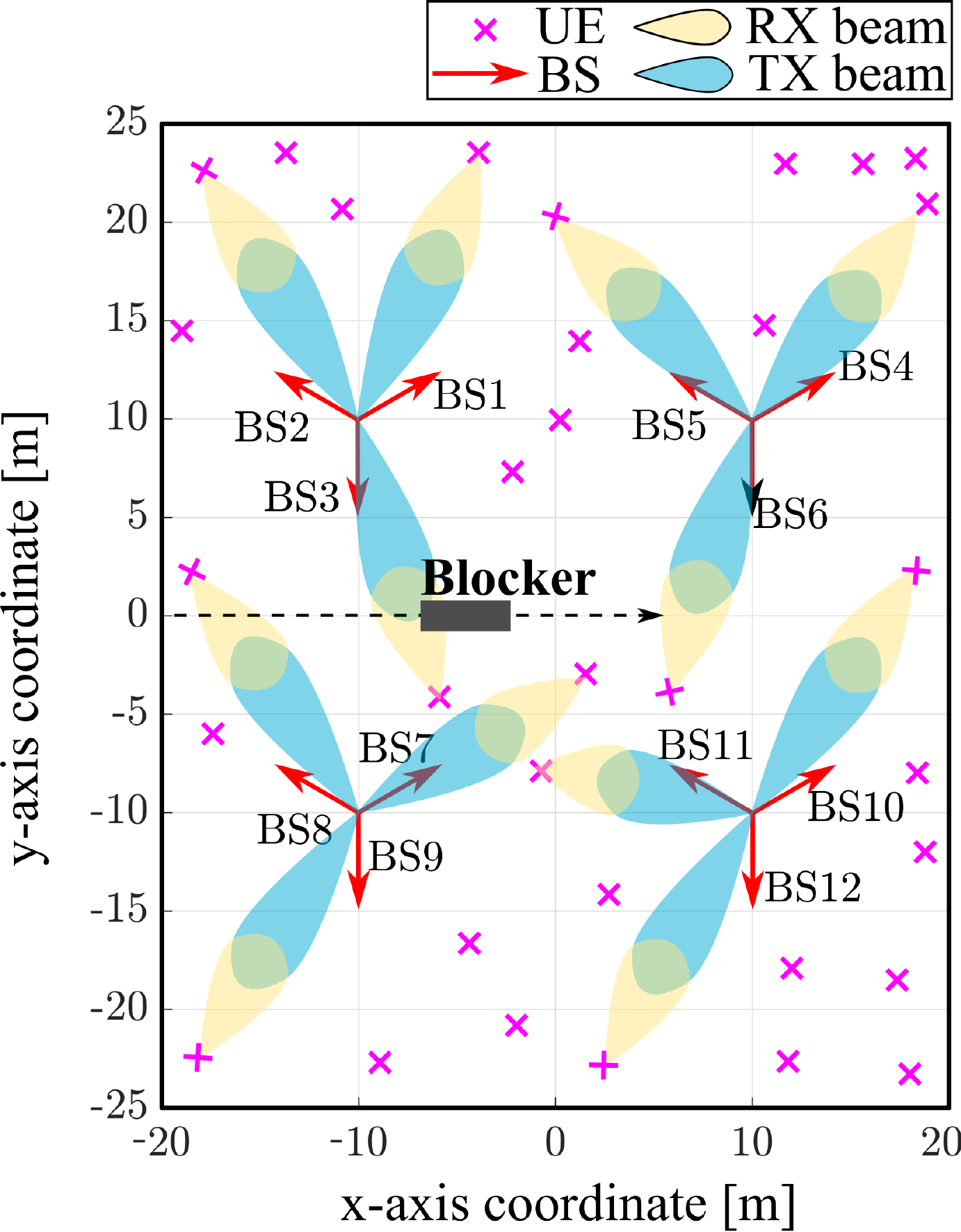}
		\caption{Layout of the indoor \ac{mmWave} network.}
		\label{fig1}
	\end{figure}
	\subsection{Network Layout}
	\label{sec:network_layout}
	We consider the network layout depicted in Fig. \ref{fig1} and inspired by the \ac{3GPP} Indoor Hot-spot scenario \cite{3gpp.38.901,3gpp.RT-170019}, formed by four cell sites deployed on a rectangular grid with an \ac{ISD} of $20$ meters. 
	Each cell site has three sectors, served respectively by three \acp{BS} oriented with an angle $\theta_j$ with $j \in \{1,2,3\}$ and operating at \ac{mmWave} carrier frequency $f_c$, with bandwidth $BW$. 
	The \acp{BS} in the \ac{mmWave} network form the set $\cJ$ with cardinality $J$. 
    We assume a set of $\cK$ \acp{UE} uniformly distributed over a rectangular area of $50$ meters by $40$ meters and we consider dynamic blockers having dimension $w \times h$ moving at speed $v$ in the environment. 
    
    \begin{figure}[t]
	\centering
    	\subfloat[Top-down view]{%
    	    \includegraphics[width=1\columnwidth]{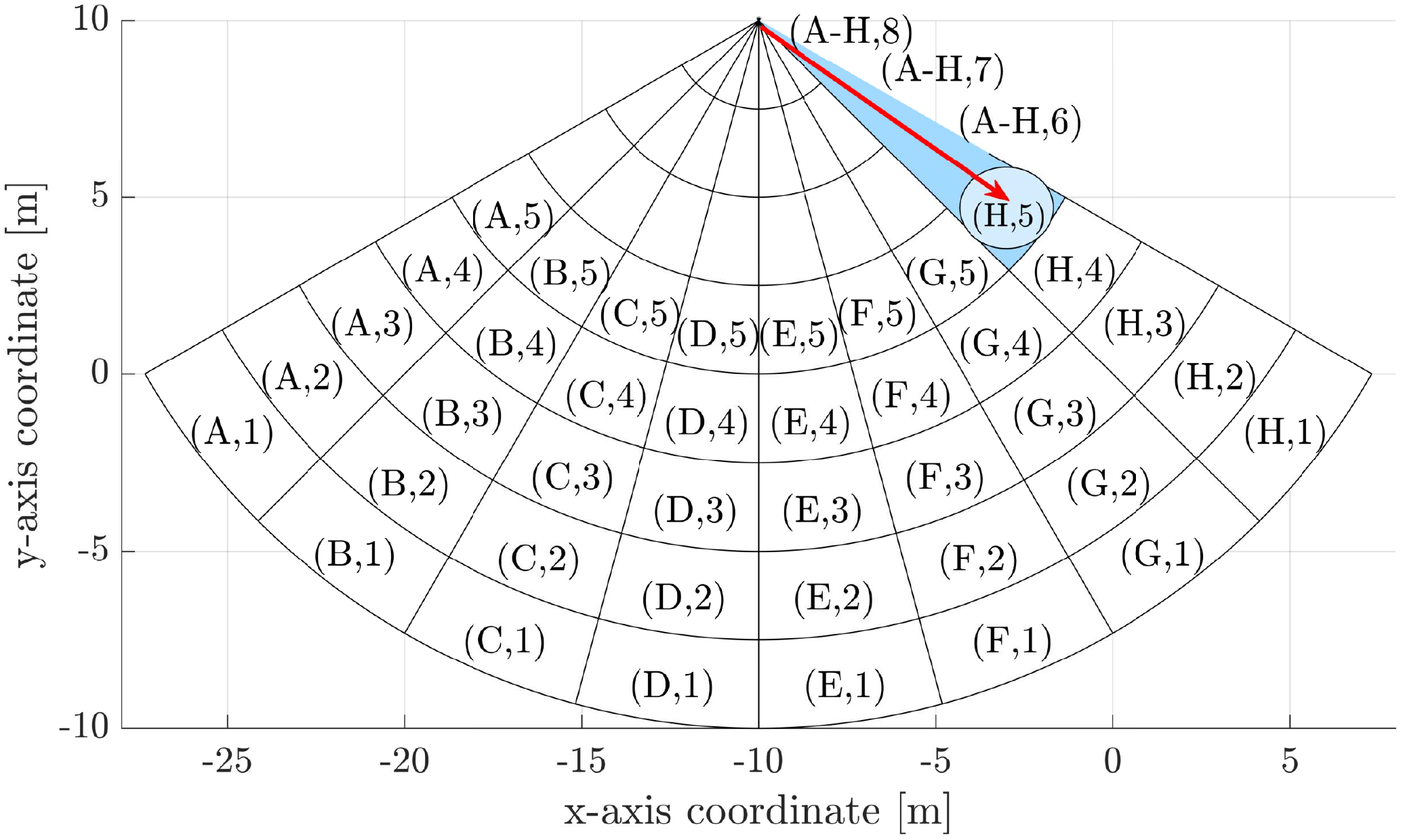}\label{fig:beamBS3Topview}
    	}
    	\vspace*{3mm}
    	\subfloat[Structure of the Tx UPA]{%
    		\includegraphics[width=0.4\columnwidth,keepaspectratio]{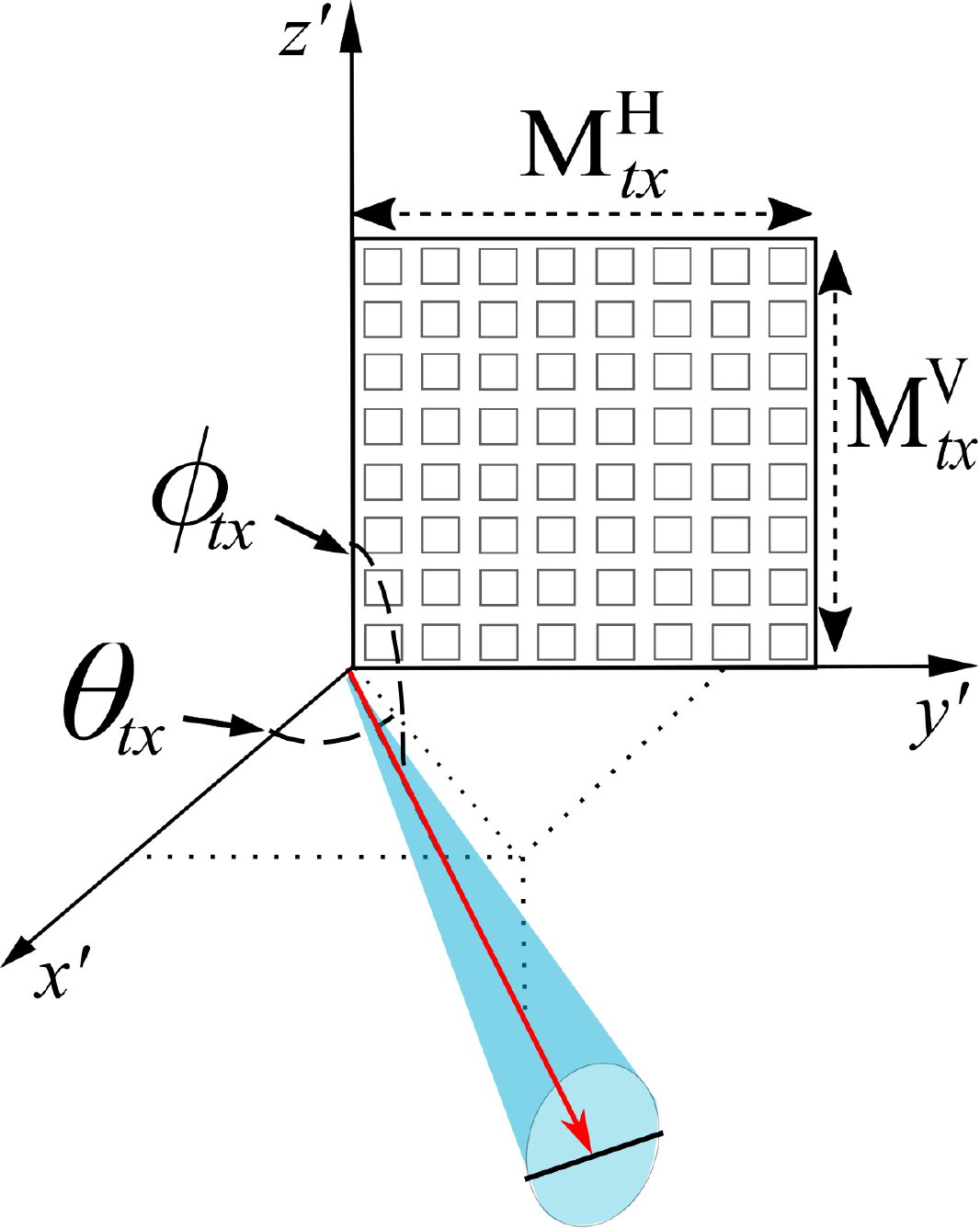}\label{fig:beamBS3_3Dview}
    	}
        \caption{(a) Top-down view of the area in the x-y plane illuminated by the Tx beams of the \ac{BS}-3. 
        (b) Structure of the UPA formed by $M_{Tx}^{\rm{H}} \times M_{Tx}^{\rm{V}}$ antennas and placed in the local coordinates system $(x',y',z')$. The Tx beam has azimuth $\theta_{Tx}$ defined between the x' axis and the Tx beam projection on the x'-y' plane and elevation $\phi_{Tx}$ defined between the z' axis and the Tx beam direction.}
    	\label{fig:beamBS3}
    \end{figure}
    
    Assuming \ac{DL} transmission, both the \ac{BS} \ac{Tx} and the \ac{UE} \ac{Rx} employ \ac{UPA} structures with $M_{Tx} = M_{Tx}^{\rm{V}} \times M_{Tx}^{\rm{H}}$ and $M_{Rx} = M_{Rx}^{\rm{V}} \times M_{Rx}^{\rm{H}}$ antennas spaced half-wavelength, where $M_{Tx}^{\rm{V}}$ and $M_{Rx}^{\rm{V}}$ are the number of \ac{Tx} and \ac{Rx} antennas in the vertical direction, and $M_{Tx}^{\rm{H}}$ and $M_{Rx}^{\rm{H}}$ are the ones in the horizontal direction. 
    We assume the \ac{BS} and \ac{UE} implement analog beamforming with a single \ac{RF} chain, selecting beams from a pre-defined codebook. 
    We denote as $(\theta_{Tx},\phi_{Tx})$ the azimuth and elevation \acp{AoD} and as $(\theta_{Rx},\phi_{Rx})$ the azimuth and elevation \acp{AoA} of the \ac{Tx} and \ac{Rx} beamforming vectors, respectively. 
    The \ac{Tx} beamforming vector for \ac{UPA} can be expressed as $\textbf{b}_{Tx}(\theta_{Tx},\phi_{Tx}) = \frac{1}{\sqrt{M_{Tx}^{\rm{H}} M_{Tx}^{\rm{V}}}} \left[1,\dots,e^{-j\pi (( M_{Tx}^{\rm{H}} -1 ) \Theta_{Tx} + ( M_{Tx}^{\rm{V}} -1 ) \Phi_{Tx})}\right]^{\sfT}$
    where 
    $\Theta_{Tx}=\sin\phi_{Tx}\sin\theta_{Tx}$ and $\Phi_{Tx}=\cos\phi_{Tx}$.
    Similarly, the \ac{Rx} beamforming vector can be given as
    $\textbf{b}_{Rx}(\theta_{Rx},\phi_{Rx}) = \frac{1}{\sqrt{M_{Rx}^{\rm{H}} M_{Rx}^{\rm{V}}}} \left[1,\dots,e^{-j\pi (( M_{Rx}^{\rm{H}} -1 ) \Theta_{Rx}) + ( M_{Rx}^{\rm{V}} -1 ) \Phi_{Rx})}\right]^{\sfT}$
    where $\Theta_{Rx}=\sin\phi_{Rx}\sin\theta_{Rx}$ and $\Phi_{Rx}=\cos\phi_{Rx}$. 
    The beamforming vectors $\textbf{b}_{Tx}$ and $\textbf{b}_{Rx}$ are identified with two beam IDs $l$ and $q$ defined as $l=\{1,\ldots,N_{\mathrm{CB},Tx}\}$ and  $q=\{1,\ldots,N_{\mathrm{CB},Rx}\}$, where $N_{\mathrm{CB},Tx}$ and $N_{\mathrm{CB},Rx}$ denote the cardinalities of \ac{Tx} and \ac{Rx} codebooks $\mathcal{B}_{Tx}$ and $\mathcal{B}_{Rx}$, respectively. 
    Fig.~\ref{fig:beamBS3Topview} shows the top-down view of the area illuminated by the Tx beams of the \ac{BS}-3. 
    Each \ac{Tx} beam of the codebook is steered towards the direction with \ac{AoD} $(\theta_{Tx},\phi_{Tx})$ as shown in Fig.~\ref{fig:beamBS3_3Dview}, and illuminates a specific area that is marked with a letter-number combination. 
    	
    \subsection{Downlink Data Transmission}
    \label{Subsec:IA}
    
    Consider the \ac{DL} of a \ac{NR}-based system for mmWave cellular communications. The \ac{NR} standard numerology employs a physical time-frequency resource division corresponding to $14$ \ac{OFDM} symbols for one \ac{TTI} in the time domain and $12$ consecutive subcarriers in the frequency domain forming a \ac{RB} \cite{3gpp.38.211}. 
    The system operates with the \ac{TDD} scheme, and each \ac{BS} uses beam sweeping to time-multiplex multiple \acp{UE}, serving one \ac{UE} at a time. 
    At a given time instant $t$, the \ac{DL} signal transmitted from the \ac{BS} $j$ to the \ac{UE} $k$ can be expressed as
    \begin{equation}
    y_{j,k}(t) = \sqrt{P_b} s_k(t) \textbf{b}_{Rx}^{\sfT} \bH_{j,k}(t) \textbf{b}_{Tx} + \textbf{b}_{Rx}^{\sfT} z_k(t),
    \label{eq:y}
    \end{equation}
    where $P_b$ is the total power of the \ac{BS}, $s_k(t)$ with $\mathbb{E}[\abs{s_k(t)}^2]=1$ is the signal transmitted, $\bH_{j,k}(t)$ represents the impulse response of the three dimensional (3D) \ac{SCM} channel (described in Appendix \ref{appendix:channelModel}), which includes the blockage loss $BL(t)$ (described in Appendix \ref{appendix:blockModel}), $z_k(t) \sim \mathcal{CN}(0,\,\sigma^{2}_z)$ is the noise seen at the $k$-th receiver and $\textbf{b}_{Tx}$ and $\textbf{b}_{Rx}$ are the \ac{Tx} and \ac{Rx} beamforming vectors used in forming the beam pair. 
    
    In the \ac{NR} standard, the initial beam pair is established during the Initial Access (IA) phase with the beam management Procedure 1 (P1), consisting of dual \ac{Tx} and \ac{Rx} beams sweeping \cite{3gpp.38.802}. 
    During this procedure, the \acp{BS} periodically broadcast \acp{SSB}, sweeping all the \ac{Tx} beams over successive \acp{SSB} and repeating the same operation $N_{\mathrm{CB},Rx}$ times. 
    On the other hand, the \ac{UE} sweeps the \ac{Rx} beam every $N_{\mathrm{CB},Tx}$ \acp{SSB}, while measuring the \ac{RSRP} for each \ac{Tx}-\ac{Rx} beam pair. 
    After the \ac{BS}-\ac{UE} sweep all the combinations of \ac{Tx} and \ac{Rx} beams, the best beam pair, called primary beam pair, is selected based on the maximum \ac{RSRP}. 
    Thus the \ac{UE} associates to the \ac{BS} providing the maximum \ac{RSRP} computed with beamforming at both \ac{Tx} and \ac{Rx} antennas \cite{3gpp.R1-1802446}. 
    The time to sweep all the \ac{Tx}-\ac{Rx} beam pairs and complete 
    an entire cycle of \acp{SSB} transmission can be expressed as\footnote{Each \acp{SSB} is mapped to $4$ \ac{OFDM} symbols of the \ac{TTI} in the time domain and $20$ \acp{RB} over $275$ \acp{RB} in the frequency domain \cite{3gpp.38.300}. 
    Multiple \acp{SSB} are grouped in a SS Burst and cover successive \acp{TTI}. 
    Multiple SS Bursts are referred to as SS Burst Set, which is transmitted in the first half-frame ($5$ ms) and has a periodicity of two \ac{NR} frames ($T_{SS}=20$ ms). 
    The maximum number of \acp{SSB} within each SS Burst Set is frequency-dependent and is equal to $L_{\rm{SSB}}=64$ for the Frequency Range 2 (FR2) \cite{3gpp.38.213}.}
    \begin{equation}
     T_{sweep} =T_{SS}\frac{N_{\mathrm{CB},Tx}N_{\mathrm{CB},Rx}}{L_{\rm{SSB}}} + \frac{T_{SS}}{2}, 
    \label{eq:Tsweep}
    \end{equation}
    where $T_{SS}/2$ is the average time to wait until the subsequent SS Burst Set transmission, assuming that the blockage happens at a time $\bar{t}\sim{\mathcal {U}}[t,t+T_{SS}]$ within the duration of two \ac{NR} frames. 

    After the beam-sweeping procedure is completed, the \ac{DL} transmission between \ac{BS} $j$ and \ac{UE} $k$ through the primary beam pair provides a data rate that can be expressed as
    \begin{equation}
    r^{'}_{j,k}(t) = \frac{BW}{K_j} \log_2\left( 1 + \frac{P_b \abs{\textbf{b}_{Rx}^{\sfT} \bH_{j,k}(t) \textbf{b}_{Tx}}^2}{I_{j,k}(t) + \sigma^{2}_z} \right),
    \label{eq:datarate}
    \end{equation}
    where $K_j$ represents the number of \acp{UE} served by the \ac{BS} $j$ and $I_{j,k}(t)$ represents the inter-cell interference, which can be expressed as 
    $I_{j,k}(t) = \sum_{j' \in \cJ \setminus j} \sqrt{P_b} s_{k'}(t) \textbf{b}_{Rx}^{\sfT} \bH_{j,k}(t) \hat{\textbf{b}}_{Tx}$, 
    where $\hat{\textbf{b}}_{Tx}$ represents the \ac{Tx} beam directed from \ac{BS} $j'$ to \ac{UE} $k'\neq k$.
	
	\section{Beam Recovery Methods}\label{Sec:3}
	
	In what follows, we consider that the primary beam pair becomes blocked when the 3D rectangular screen modelling the blocker intersects the \ac{LoS} path between \ac{BS} $j$ and \ac{UE} $k$.\footnote{This assumption is primarily motivated by the indoor scenario, where we measured a probability of having a \ac{LoS} link between the \ac{BS} server and \ac{UE} that is $99.9\%$.} 
    At time $t$, the \ac{GT} state of the \ac{Tx} beam $l$, forming the primary beam pair, can be expressed as
    \begin{equation}
    S_{l}(t) = 
    \begin{dcases}
    0 \quad \text{(non-blocked)} & \text{for $0 < t < \bar{t} \, \cup \, t \geq  \bar{t}+T_{bl}$,} \\
    1 \quad \text{(blocked)} & \text{for $\bar{t} \leq t <  \bar{t} + T_{bl}$,}
    \end{dcases}
    \label{eq:blstate}
    \end{equation}
    where $T_{bl}$ represents the blockage event duration.
    
    In the \ac{NR} standard, the status of the beam is monitored through the \ac{RSRP} measurements, based on the Channel State Information - Reference Signals (CSI-RSs) received at the \ac{UE}, and reported periodically back to the \ac{BS} using the \ac{UL} control channels \cite{3gpp.38.213,3gpp.38.300}. 
    The \ac{RSRP} values can be mapped to the $\snr$ through a linear relationship.\footnote{The mapping between \ac{RSRP} and $\snr$ is valid under the assumption that the \ac{RSRP} does not account for the inter-cell interference, since it is computed with the average power of all the cell-specific CSI-RSs carried over multiple \acp{RB}.} 
    For instance, $\snr[\rm{dB}] = \rm{RSRP}[\rm{dBm}]+122~\rm{dBm}$, when the receiver noise is $-122~\rm{dBm}$.
    Thus, we consider to track the quality of the primary beam pair with the instantaneous $\snr$, expressed as $\snr_l(t) = \frac{ P_b \abs{\textbf{b}_{Rx}^{\sfT} \bH(t) \textbf{b}_{Tx}}^2} {\sigma^{2}_z}$ and the average value of $\snr$ for the non-blocked time instants denoted as $\overline{\snr}_l$. 
    
    When the primary link becomes blocked, 
    a neighbour \ac{BS} in proximity of the coverage area of the server \ac{BS}, can provide a backup beam pair from a different spatial direction of the primary beam pair and most likely not affected by the blockage when the primary is blocked, as shown in Fig.~\ref{fig:ch4BRDetMethodSpace}. 
    Hence, we consider that the \ac{UE} identifies a secondary \ac{BS}, for instance, during the IA phase, which provides the second-largest \ac{RSRP} computed adopting beamforming at both \ac{Tx} and \ac{Rx} antennas. 
    We assume to repeat the beam sweeping procedure to identify the backup beam pair at every blockage event since, in dynamic environments, the beam training results may not stay the same for a long time \cite{6134444}. 
    Additionally, we assume that data transmission occurs during the beam sweeping procedure $T_{sweep}$. 
    This is because, as described before in Sec.~\ref{Subsec:IA}, the \acp{SSB} use limited spectrum resources, i.e. 20 \acp{RB} over specific \ac{OFDM} symbols, and the remaining spectrum resources may be dedicated to data transmission \cite{8458146}.\footnote{The data transmission in the \ac{SSB} \acp{TTI} is possible under certain circumstances. 
    For instance, the secondary \acp{BS} may serve those \acp{UE} in the same direction of the \ac{SSB} transmission \cite{8458146}, whereas the serving \acp{BS} may schedule the \ac{UE} in different \acp{TTI} from the ones used by the \acp{SSB} or the \ac{UE} may perform beam sweeping within the \ac{TTI} used for \ac{SSB} transmissions.}
	
	\subsection{Beam Recovery Based on Blockage Detection}
	\label{sec:blockage detection}
	
	\begin{figure*}[t]
	\centering
    	\subfloat[Dynamic blockage scenario with primary and backup beams represented for one of the \ac{UE}.]{%
    		\includegraphics[width=0.7\columnwidth]{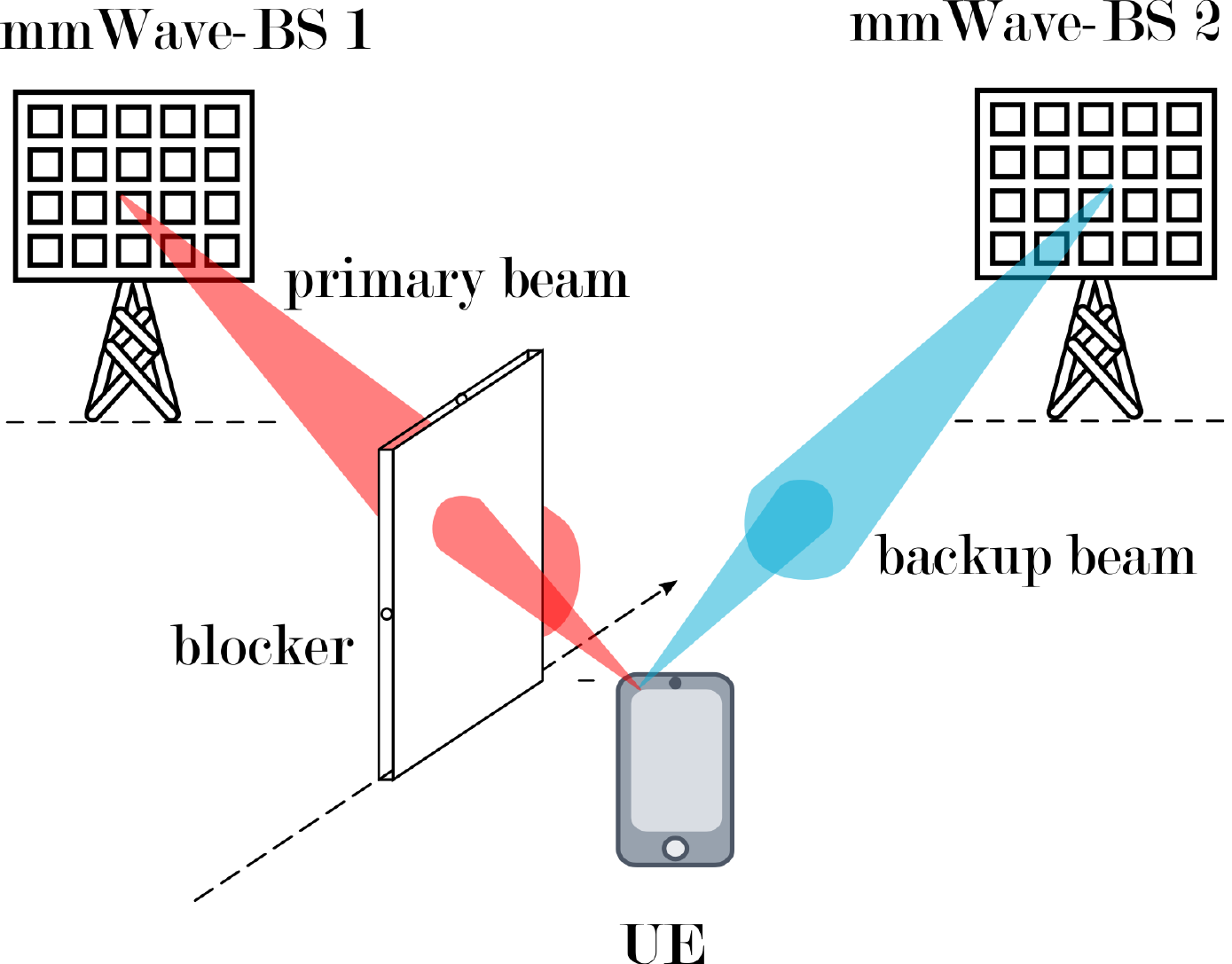}\label{fig:ch4BRDetMethodSpace}
    	}
    	\hfil
    	\subfloat[Temporal variations of the SNR for primary and backup beams during the blockage event.]{%
    		\includegraphics[width=0.8\columnwidth]{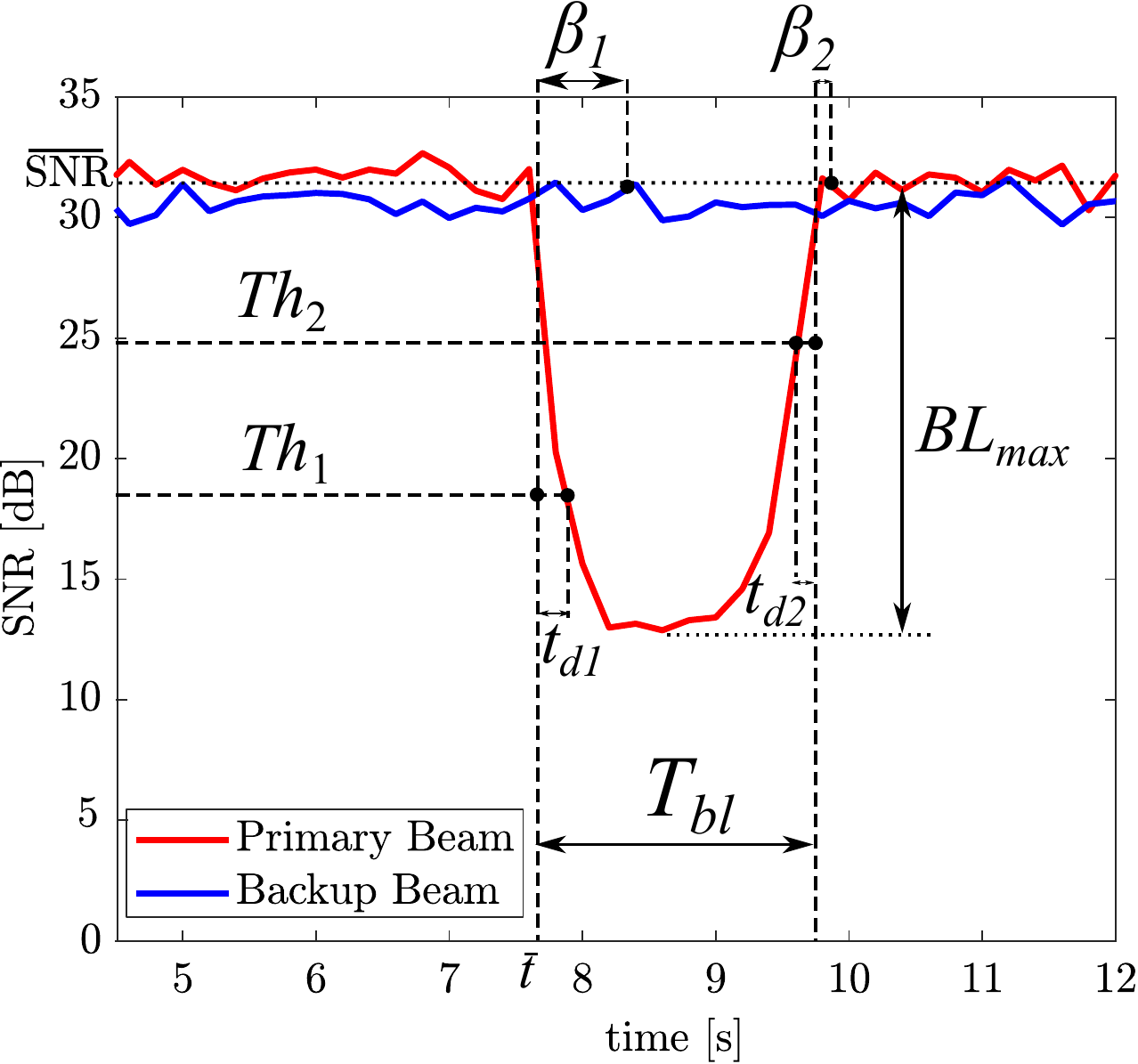}\label{fig:ch4BRDetMethod}
    	}
    	\caption{Beam switching operations of the method based on detection according to the temporal variations of the SNR values for primary and backup beams.}
    \end{figure*}
    
	In this section, we describe a potential implementation of the current state-of-art method based on a detection threshold to recover the beam when the \ac{mmWave} links are affected by blockage events, as represented in Fig~\ref{fig:ch4BRDetMethodSpace}. 
    The beam recovery based on the detection (BR-Det) method is considered the benchmark for our studies and operates in three phases. 
    Firstly, it identifies the blockage events when the variations of \ac{DL} $\snr$ surpass a threshold. Then, it recovers the blocked beams after aligning the backup beam pair with the secondary \ac{BS}, and finally, it switches back to the primary beam once the blockage is cleared. 
    We define as $\mathit{Th}_1$ and $\mathit{Th}_2$ two thresholds with 
    values corresponding to $\overline{\snr}_l$ minus $70\%$ and $30\%$ of the maximum blockage loss $BL_{max}$, respectively, in line with the procedure adopted in \cite{8895815}. 
    
    As shown in Fig.~\ref{fig:ch4BRDetMethod}, whenever the beam is non-blocked, the $\snr$ shows small fluctuations around the average value $\overline{\snr}_l$ due to the channel's multipath components. 
    At the time instant $\bar{t}$ when the blocker starts to intersect the primary beam, the $\snr$ starts to decay rapidly, and the blockage event is detected if $\snr_l(t) < \mathit{Th}_1$. 
    The beam sweeping procedure follows this event to align the backup beam pair. 
    The time that it takes to handover the communication to the secondary \ac{BS} from the time when the blockage begins can be expressed as 
    \begin{equation}
    \beta_1= t_{d1} + T_{sweep} + T_{HO},
    \label{eq:beta1}
    \end{equation}
    where 
    $t_{d1}$ is the interval between $\bar{t}$ and the time instant when the $\snr$ decays below the threshold $\mathit{Th}_1$, $T_{sweep}$ defined in Eq.~\eqref{eq:Tsweep} denotes the beam sweeping duration to identify the backup beam pair and $T_{HO}$ represents the time interval before the handover to the secondary \ac{BS}. 
    After switching to the backup beam pair, the data rate of the \ac{DL} transmission between the secondary \ac{BS} $j'$ and \ac{UE} $k$ can be expressed as
    \begin{equation}
    r^{''}_{j',k}(t) = \frac{BW}{K_{j'}} \log_2\left( 1 + \frac{P_b \abs{\bar{\textbf{b}}_{Rx}^{\sfT} \bH_{j',k}(t) \bar{\textbf{b}}_{Tx}}^2}{I_{j',k}(t) + \sigma^{2}_z}\right),
    \label{eq:datarate2}
    \end{equation}
    where $\bar{\textbf{b}}_{Tx}$ and $\bar{\textbf{b}}_{Rx}$ denote the \ac{Tx} and \ac{Rx} beamforming vectors of the backup beam pair.
    
    At the end of the blockage event, which lasts for the duration $T_{bl}$, the primary beam $\snr$ returns to values without blockage, i.e. $\overline{\snr}_l$, and the \ac{UE} switches back to the primary beam pair. 
    The time that it takes to handover the communication back to the \ac{BS} $j$ from the time when the blockage ends can be expressed as 
    \begin{equation}
    \beta_2= T_{HO} - t_{d2},
    \end{equation}
    where $t_{d2}$ is the interval between the time instant $\bar{t} + T_{bl}$, representing the end of the blockage and the time instant when $\snr(t) > \mathit{Th}_2$, and $T_{HO}$ is the time before the handover to \ac{BS} $j$. 
    
    The data rate of \ac{UE} $k$ obtained while adopting the BR-Det method can be expressed as
    \begin{equation}
    r^{\mathrm{BRDet}}_{k}(t) = 
    \begin{dcases}
    r^{'}_{j,k}(t)
    \qquad\text{for $0 < t < \bar{t} + \beta_{1}$,}
    \\
    r^{''}_{j',k}(t)
    \qquad\text{for $\bar{t} + \beta_{1} \leq t < \bar{t} + T_{bl} + \beta_{2}$,}
    \\
    r^{'}_{j,k}(t)
    \qquad\text{for $t  \geq \bar{t} + T_{bl} + \beta_{2}$.}
    \end{dcases}
    \label{eq:datarate_withrecovery}
    \end{equation}
		
	\subsection{Beam Recovery Based on Blockage Predictions}\label{sec:blockage prediction}
	
	In this section, we propose a novel beam recovery method based on predictions (BR-Pre) capable of switching to the backup beam pair, before the blocker affects the quality of the ongoing communication over the primary beam pair. 
    
    Firstly, we define the prediction window $\eta$ to be larger than the beam sweeping duration $T_{sweep}$ and the handover period $T_{HO}$, i.e. $\eta>T_{sweep}+T_{HO}$. 
    The BR-Pre method evaluates at the beginning of the prediction window, i.e. at the time $t-\eta$, the beam state predictions $\hat{S}_l$ indicating whether the blockage is going to obstruct the primary beam after the end of the prediction window $\eta$, at time $t$. 
    If the beam is predicted as non-blocked, i.e. $\hat{S}_l(t)=0$, the BR-Pre method continues using the primary beam pair. 
    Differently, when the beam is predicted as blocked, i.e. $\hat{S}_l(t)=1$, the BR-Pre method uses this early indication to start the beam sweeping procedure in advance to establish the backup beam pair.\footnote{We recall that the beam sweeping procedure can be performed during data transmission as it only uses a partial set of \acp{RB} and \ac{OFDM} symbols in the \acp{TTI} where the \ac{SSB} transmission occurs.}
    
    At the end of the prediction window, the BR-Pre method switches to a newly computed backup beam pair and the data rate of the \ac{DL} transmission between the secondary \ac{BS} $j'$ and \ac{UE} $k$ can be expressed as in Eq.~\ref{eq:datarate2}. 
    We want to emphasise that the advantage of the BR-Pre against the BR-Det method is that at the time of the beam switching (assuming the beam state predictions are correct, i.e. $\hat{S}_l(t)={S}_l(t)$) the primary beam $\snr$ has not yet dropped since the beam switching happens at the end of the prediction window and before the blockage begins, while with the BR-Det method described in Sec.~\ref{sec:blockage detection}, the beam switching happens with a delay $\beta_1$ after the blockage begins. 
    
    During the blockage, the BR-Pre method continues to evaluate the predictions of the beam $l$, and if the state becomes non-blocked, i.e. $\hat{S}_l(t)=0$, the BR-Pre switches back to the primary beam pair. 
    The data rate of \ac{UE} $k$ obtained while adopting the BR-Pre method can be expressed as
    \begin{equation}
    r^{\mathrm{BRPre}}_{k}(t) = 
    \begin{dcases}
    r^{'}_{j,k}(t) \qquad\text{if $\hat{S}_l(t)=0$,}
    \\
    r^{''}_{j',k}(t) \qquad\text{if $\hat{S}_l(t)=1$.}
    \end{dcases}
    \label{eq:datarate_withprediction}
    \end{equation}

	\section{Beam State Predictions}
	\label{Sec:4}

	\begin{figure}[t]
		\centering
		\includegraphics[width=\columnwidth,height=4.5cm,keepaspectratio]{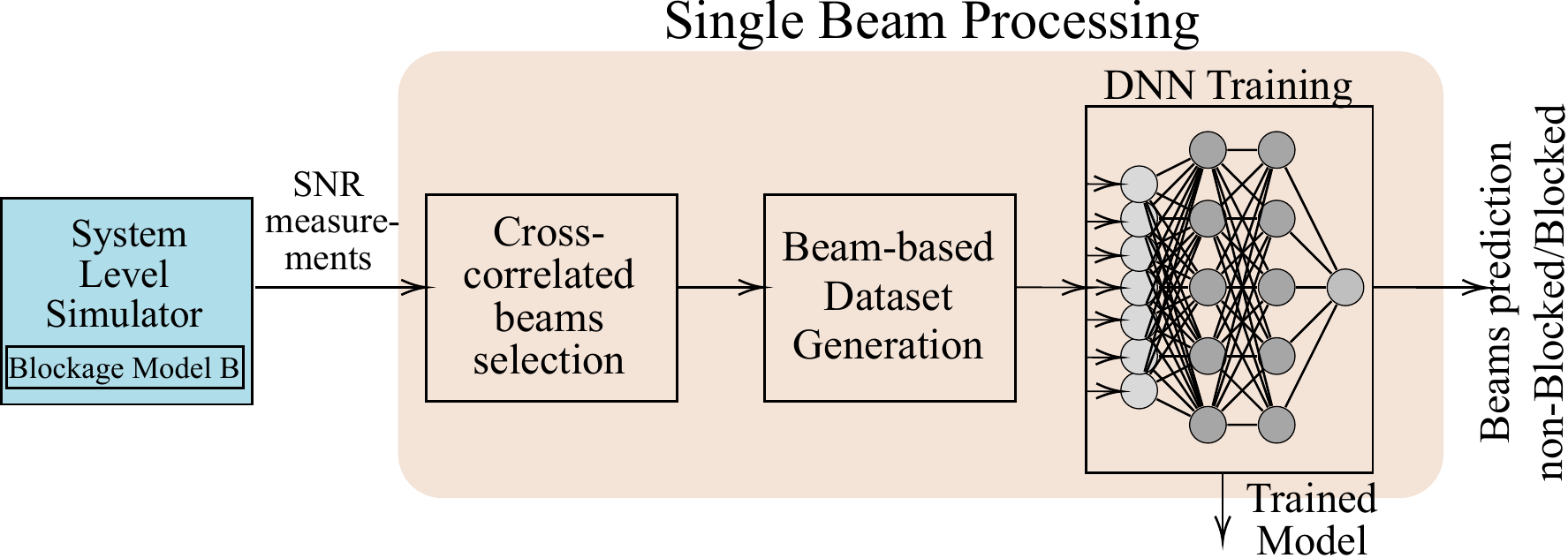}
		\caption{Key steps of the procedure used to obtain the beam state predictions.}
		\label{fig:syst_arch}
	\end{figure}
	
	In this section, we propose a procedure to obtain the beam state predictions $\hat{S}_l$ adopted in the BR-Pre method. 
    The main idea is to use the $\snr$ measurements of other \ac{Tx}-\ac{Rx} beam pairs -- exchanged between \acp{BS} through a central server -- to predict $\eta$ time instants before the blockage changes the state of the beam $l$ at time $t$. 
    As the blockage effect follows in general non-periodic temporal dynamics, the $\snr$ does not change until the blocker intersects the beam $l$, making it unpractical to infer the beam-state only by analysing the $\snr$ temporal variations of the single beam. 
    On the other hand, by looking at the $\snr$ temporal variations of other beams, their $\snr$ may drop earlier than the $\snr$ of the beam to predict. 
    We train a \ac{DNN} model to relate these early $\snr$ variations of other beams (inputs) to the blockage states of the beam $l$ in successive instants of time (output). 
    
    Furthermore, as the \ac{BS} is adopting a codebook with multiple beams, the \ac{DNN} model needs to simultaneously provide multiple beam state predictions. 
    For example, with a codebook formed by two beams, the first label indicates the state of the first beam (blocked, non-blocked) while the second label indicates the state of the second beam, assumed independent of the state of the first beam. 
    Therefore, the task consists of learning one binary classifier per beam, meaning that we train and validate a beam-specific \ac{DNN} model for each beam. 
    As shown in Fig.~\ref{fig:syst_arch}, a sequence of steps leads to the beam-specific prediction $\hat{S}_l$. 
    We describe each step in the sections that follow. 
	
    \subsection{Cross-correlated Beams Selection}
    
    In this subsection, we introduce a pre-processing step that extracts a subset of beams to use for the predictions considering the cross-correlation between the $\snr$ time series of different beams. 
    This procedure is motivated by two observations. 
    First, the network presents many beams that will be irrelevant for predicting a specific beam state. 
    This could needlessly complicate the training process without delivering any improvement. 
    Second, since correlated beams are most likely spatially close, they also belong to the same \ac{BS} or neighbour \acp{BS}. 
    Therefore, we can also optimise the exchange of information among \acp{BS} to limit the use of a central server. 

    Let us consider the pair of \ac{Tx} beams $l$-$l'$. 
    The delay between $l$ and $l'$ is estimated as $\delta_{l,l'} = \argmax_{\forall \tau} R_{l, l'}(\tau)$, where $R_{l, l'}(\tau) = \sum_{t = 0}^{T} \snr_{l}(t) \snr_{l'}(t - \tau)$ is the cross-correlation between the $\snr$ time series of $l$ and the $\snr$ time series of $l'$ defined over the time window $[0,T]$, where $T$ is the total observation time. 
    We estimate the delays between $l$ and all the other beams $l'\in \cB~\setminus~l$, where $\cB = \{1,2,\ldots,(N_{\mathrm{CB},Tx}\times J)\}$ represents the set of all the network's beams and $B=N_{\mathrm{CB},Tx}\times J$ is the total number of beams in the network. 
    Then, we pick the $L$ beams that have the lowest delays in absolute value, indicating that they are spatially close to beam $l$ and we define a set $\cC_l=\{c_1, c_2,\ldots,c_{L}\}$, which contains the beams to use for predicting the state of beam $l$. 
    In Algorithm \ref{Algo1}, we describe the procedure in details.
    
    \begin{algorithm}[t]
    	\SetAlgoLined
    	\KwIn{A set $\cS = \{\snr_{1}(t), \ldots, \snr_B(t)\}$ of time series defined in $0 < t < T$}
    	\KwOut{Set of cross-correlated beams $\cC_l$}
    	$\var{DelaysAll} \gets$ Vector(size: $B-1$)\;
    	$\var{BeamsAll} \gets$ Vector(size: $B-1$)\;
    	$\cC_l \gets$ Vector(size: $L$)\;
    	\For{$l' \gets 1$ \textbf{to} $B-1$} {
    		\For{$\tau \gets 0$ \textbf{to} $T$} {
    			$R_{l, l'}(\tau) \gets $ Compute cross-correlation\;
    		}
    		$\delta = \argmax_{\forall \tau} R_{l, l'}(\tau)$ \tcp*[l]{Delay $l$-$l'$}
    		$\var{DelaysAll} \gets \var{DelaysAll} \cup \{\delta\}$\; 
    		$\var{BeamsAll} \gets \var{BeamsAll} \cup \{l'\}$\; 
    	}
    	Sort $\var{BeamsAll}$ by the absolute value of $\var{DelayAll}$\;
    	Copy the first $L$ entries of $\var{BeamsAll}$ into $\cC_l$\;
    	\Return $\cC_l$\;
    	\caption{Correlated beams selection for beam $l$}
    	\label{Algo1}
    \end{algorithm}
   	
   	\subsection{Dataset Generation}
    \label{sec:dataset_generation} 

	\begin{figure*}[t]
    	\centering
    	\subfloat[Example of input data corresponding to the non-blocked output state.]{\includegraphics[width=0.9\textwidth]{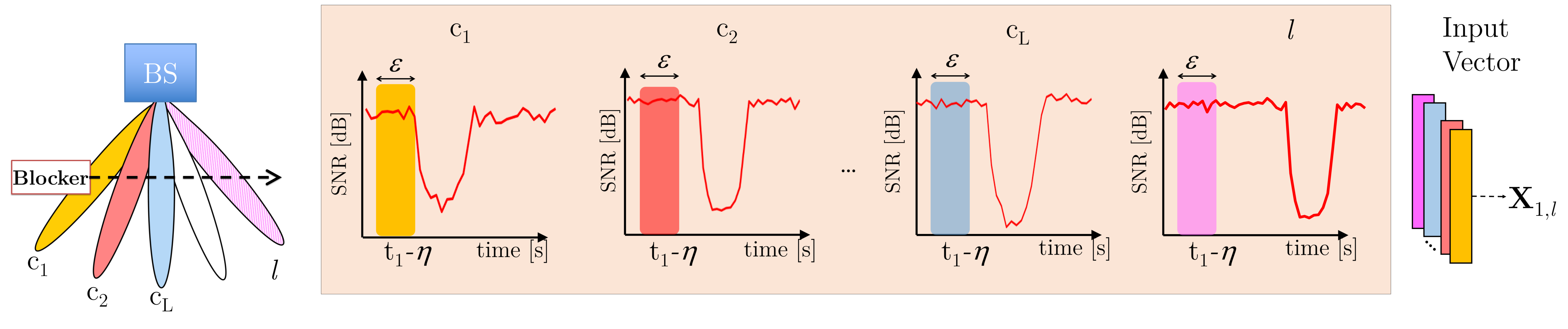}\label{fig:dataMappingInputNonBlocked}}%
    	\\
    	\vspace*{3mm}
    	\subfloat[Example of input data corresponding to the blocked output state.]{\includegraphics[width=0.9\textwidth]{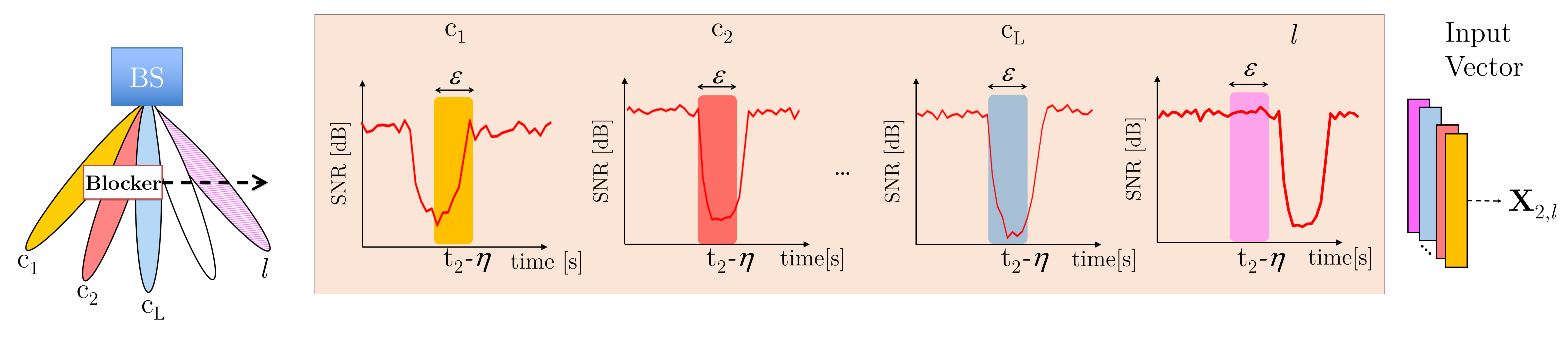}\label{fig:dataMappingInputBlocked}}%
    	\\
    	\vspace*{3mm}
    	\subfloat[Examples of output data for non-blocked and blocked states of the magenta beam.]{\includegraphics[width=0.45\textwidth]{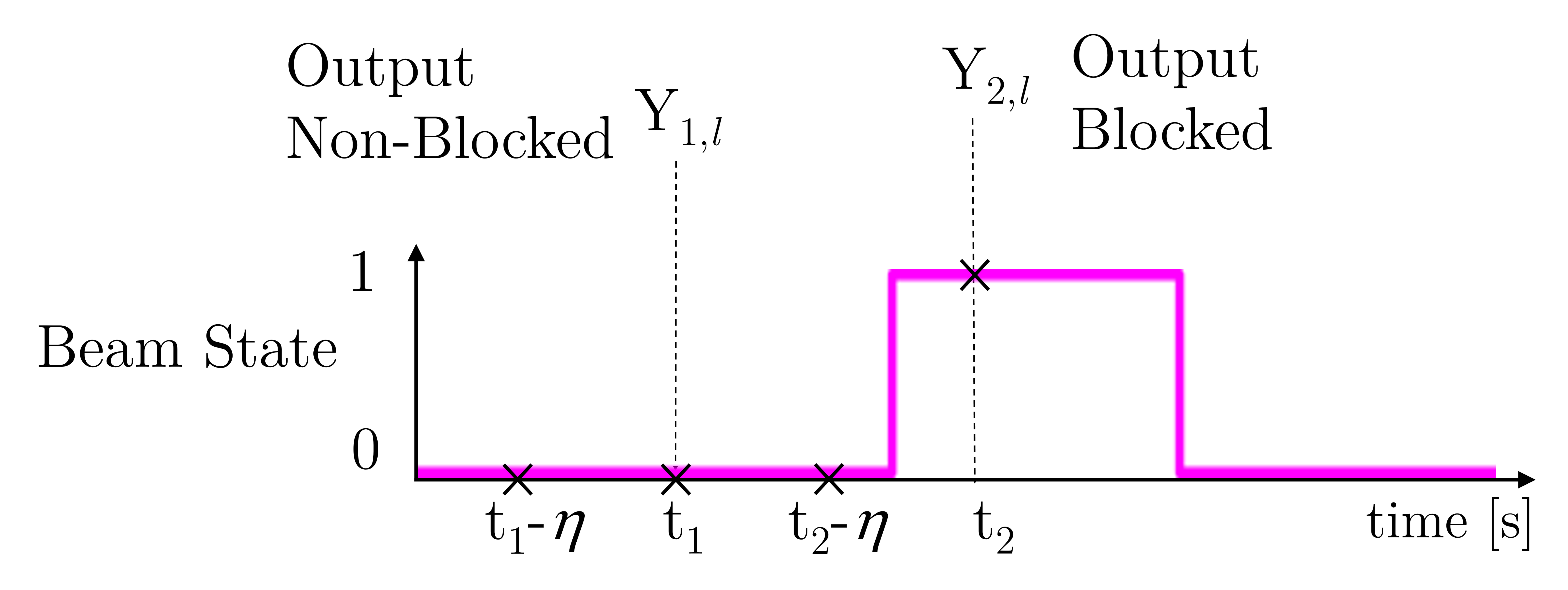}\label{fig:dataMappingOutput}}%
    	\caption{Two examples of input-output vectors forming the dataset for the beam depicted with the line pattern (magenta) on the left side of Figs \ref{fig:dataMappingInputNonBlocked} and \ref{fig:dataMappingInputBlocked}. 
    	In Fig.~\ref{fig:dataMappingInputNonBlocked}, the input data are taken at a time $t_1-\eta$ from the $\snr$ measurements and correspond to the non-blocked state at the time $t_1$ in Fig.~\ref{fig:dataMappingOutput} . 
    	Conversely, in Fig.~\ref{fig:dataMappingInputBlocked}, 
    	the input data are taken at a time $t_2-\eta$ and correspond to the blocked state at the time $t_2$ in Fig.~\ref{fig:dataMappingOutput}.}
    	\label{fig:dataMapping}
    \end{figure*}

    We consider a dataset formed by $N$ samples, where the $n$-th sample of the dataset is organised into input/output vectors generated as follows: 
    \begin{itemize}
        \item \textbf{Input:} let us define as $\bv_{l}= [\snr_{l} [t-\eta-(\epsilon-1)],\snr_{l} [t-\eta-(\epsilon-2)],\ldots,\snr_{l} [t-\eta]]^{\sfT}$ the vector containing $\epsilon$ samples of the $\snr$ for beam $l$. We assume that the $\snr$ measurements from the beams in the set $C_l=\{c_1, c_2,\ldots,c_{L}\}$ are exchanged with the \acp{BS} through a central server. 
        These data are also represented for each beam $l' \in \cC_l$ with a vector $\bv_{l'}= [\snr_{l'} [t-\eta-(\epsilon-1)],\snr_{l'} [t-\eta-(\epsilon-2)],\ldots,\snr_{l'}[t-\eta]]^{\sfT}$ containing $\epsilon$ samples of the $\snr$. 
        Thus, the input is defined as $\bX_{n,l}=\left[\bv_l,\bv_{c_1},\bv_{c_2},\ldots,\bv_{c_{L}}\right] \in \bbR^{ \epsilon \times (L+1)}$ and aggregates the $\snr$ measurements of $L+1$ beams altogether.
        
        \item \textbf{Output:} the output is constituted by a variable $Y_{n,l}$ set to $1$ (blocked) or $0$ (non-blocked) according to the \ac{GT} state of the beam $l$ at time instant $t$. 
        We obtain this information from the simulation environment. We set the label to $1$ (blocked), when the blocker intersects the \ac{LoS} path between \ac{BS} and \ac{UE}. 
        Conversely, we set the label to $0$ (non-blocked) when there is no intersection. 
    \end{itemize}
    If the beam is assigned to multiple \acp{UE}, we fill in the input vector with the median $\snr$ and output variable corresponding to the most frequent label inside the \acp{UE} group. 
    If there are no \acp{UE} assigned to a beam, the corresponding $\snr$ measurement cannot be recorded, and the input's entries are filled with a constant $\snr$ (in our simulations, we use $60$ dB) outside the range of the possible $\snr$ values. 
    
    Figs. \ref{fig:dataMappingInputNonBlocked} and \ref{fig:dataMappingInputBlocked} show a simplified scenario where a set of beams shown on the left side of the \ac{BS} is used to predict the state of the beam $l$, depicted with the line pattern (magenta) on the right side of the \ac{BS}. 
    The entries of the dataset are formed by associating the inputs $\bX_{1,l}$ and $\bX_{2,l}$ to the values of the \ac{GT} state $Y_{1,l}$ and $Y_{2,l}$ reported in Fig.~\ref{fig:dataMappingOutput}. 
    The two different examples show how the input data changes depending on the value of the ground-truth. 
    In the first example, depicted in Fig.~\ref{fig:dataMappingInputNonBlocked}, the $\snr$ measurements (within the sliding window of length $\epsilon$) show small variations indicating that the set of beams at the time $t_1-\eta$ is not affected by the blocker. 
    This behaviour suggests that also the magenta beam is likely not affected by the blocker at the instant $t_1$. 
    Indeed, as shown in Fig.~\ref{fig:dataMappingOutput}, the \ac{GT} state corresponds to the non-blocked value, i.e. $Y_{1,l}=0$. 
    On the other hand, in the second example, depicted in Fig.~\ref{fig:dataMappingInputBlocked}, the $\snr$ measurements show large drops for the set of beams at the time $t_2-\eta$, indicating that the magenta beam is likely affected by the blocker movement at the time $t_2$. 
    This is confirmed by the \ac{GT} state $Y_{2,l}=1$, which corresponds to the blocked value.
   	
    \subsection{Deep Neural Network Model Training}
    \label{section:dnn}

    \begin{table}[t]
    \centering
    \caption{ \ac{DNN} network parameters}
    \begin{tabular}{|c|c|c|}
    \hline
    \multirow{6}{*}{\begin{tabular}[c]{@{}c@{}} Network \\ architecture \end{tabular}} & Number of layers & $2$ (Fully-connected) \\ \cline{2-3} 
     & Number of neurons & $20$ \\ \cline{2-3} 
     & Weights initialisation & He \cite{He_2015_ICCV} \\ \cline{2-3} 
     & Activation function & ReLu \cite{2018arXiv180308375A} \\ \cline{2-3} 
     & Dropout & None \\ \cline{2-3} 
     & Normalisation Input layer & ``Zero--center'' \\ \hline
    \multirow{5}{*}{\begin{tabular}[c]{@{}c@{}} Training \\ options \end{tabular}} & Solver name & Adam \cite{2014arXiv1412.6980K} \\ \cline{2-3} 
     & Mini-batch size & $1000$ \\ \cline{2-3} 
     & Number of Epochs & $50$ \\ \cline{2-3} 
     & Learning rate & $10^{-3}$ \\ \cline{2-3} 
     & L2 regularisation & $10^{-5}$ \\ \hline
    \end{tabular}
    \label{table:DNN_parameter}
    \end{table}
    
    For training the \ac{ML} model, we employ a \ac{DNN} structure constituted by: \emph{i)} an input layer with $\epsilon\times(L+1)$ inputs, \emph{ii)} two hidden layers, and \emph{iii)} a $\var{softmax}$ function as output layer. 
    We train each \ac{DNN} model with \ac{Adam} optimiser adopting He weights initialisation and \ac{ReLu} activation functions \cite{He_2015_ICCV,2018arXiv180308375A}. 
    
    During the training, we divide the dataset into mini-batch of size $N_{bs}$ assuming a constant learning rate. 
    We repeat the training for several epochs to find the set of model parameters denoted as $\bOmega_l$ that minimises the cross-entropy loss, given the set of inputs $\bX_{1,l}, \ldots, \bX_{N_{bs},l}$ and the set of \ac{GT} labels $Y_{1,l},\ldots, Y_{N_{bs},l}$. 
    The lower is the loss of the model output compared to the ground-truth, and the better is the model. 
    
    Moreover, since the time spent by the beam in the non-blocked state prevails over the time spent in the blocked state, the dataset presents a skewed distribution towards the non-blocked class that implies a class imbalance problem \cite{5128907}. To re-balance the loss, we assign a higher weight to the cross-entropy loss computed for the rare class (blocked) and a lower weight to the cross-entropy loss computed for the dominant class (non-blocked). 
    The weighted formulation of the cross-entropy loss can be expressed as
	\begin{equation}
	\begin{split}
	\mathcal{L}_{l}(\bOmega) = -\frac{1}{N_{bs}}\sum_{n=1}^{N_{bs}}
	&\mu_{1,l}Y_{n,l}\log(\widehat{Z}_{n,l}( \bOmega))\\
	&+\mu_{2,l}(1-Y_{n,l}) \log(1-\widehat{Z}_{n,l}( \bOmega)),
	\end{split}
	\label{eq:lossfunction}
	\end{equation}
	where $\widehat{Z}_{n,l}$ is the output of the $\var{softmax}$ function and indicates the likelihood for the beam $l$ to be in the state non-blocked or blocked. Moreover, the class weights $\mu_{1,l}$ and $\mu_{2,l}$ are computed as the inverse of the number of training samples for each class, i.e. $\mu_{1,l} = N/{\sum_{n=1}^{N} \left[Y_{n,l}=0\right]}$ and $\mu_{2,l} = N/{\sum_{n=1}^{N} \left[{Y_{n,l}=1}\right]}$, respectively.
	
	Table \ref{table:DNN_parameter} reports the training configurations and the list of hyper-parameters such as the number of hidden neurons, learning rate and L2 regularisation with the initial configuration values we adopted for the training. 
    Note that the L2 regularisation incorporates in the loss function depicted in Eq.~\eqref{eq:lossfunction} a penalty term on the neuron weights, which reduces the overfit to the training dataset and improves the generalisation of the \ac{DNN} model \cite{Bishop:2006}.
	
	\section{Evaluation}\label{Sec:5}
	
	In this section, we first describe the setup used for generating the synthetic data and preparing the dataset for the training. 
    Then, we validate the results of the \ac{DNN} model state predictions for multiple beams using a training dataset. 
    Finally, we deploy the \ac{DNN} models online into the simulator to study the performance of our proposed BR-Pre method and compare with the performance of the BR-Det method, which we consider as our benchmark. 

	\subsection{Experiment Setup}
	
	The preparation of the dataset used for training the \ac{DNN} models involves several steps that entail the setup of the \ac{3GPP}-based system-level simulator, including the setup of the blocker's movement, the generation of a beam-specific dataset, the selection of the validation metric and the configuration of the prediction parameters. 
	Now we discuss each of these steps:
	
	\emph{1) Simulation setup:} we take a drop-based approach to generate synthetic data utilising a detailed and calibrated system-level simulator that implements \ac{3GPP} 3D \ac{SCM} for the 0.5-100 GHz spectrum \cite{3gpp.38.901}. 
	We configure the simulator according to the list of parameters reported in Table \ref{table:system_level_parameters} of Appendix \ref{appendix:simulationParameters}. 
	We simulate the \ac{mmWave} network shown in Fig.~\ref{fig1}, placing the \acp{BS} on a rectangular grid and the \acp{UE} in random positions, which are kept fixed between drops. 
	We assume time-invariant large-scale channel parameters (LSPs) such as delay spread, \ac{AoD} and \ac{AoA} spread, Rician K-factor, shadow fading (SF) and pathloss, for the duration of the drop $T$. 
	Additionally, at each simulation step within a drop, we assume \emph{i)} time-variant blockage loss $BL(t)$ due to the blocker's movement and \emph{ii)} time-variant small-scale channel parameters (SSPs), e.g. power, phase, delay, \ac{AoA} and \ac{AoD} for each \ac{MPC}, that change independently between simulation steps due to the variations of the multipath fading. 
	We fix the network drop duration to $T=40$ s, and we consider an interval $\Delta t=200$ ms to update the channel $\bH$ between the simulation steps. 
	We compute every \ac{BS}-\ac{UE} link's channel and the \ac{DL} $\snr$ for each beam and for all the \acp{RB} of the system bandwidth at each simulation step. 
	\begin{figure*}[t]
    	\centering
    	\includegraphics[width=0.8\textwidth]{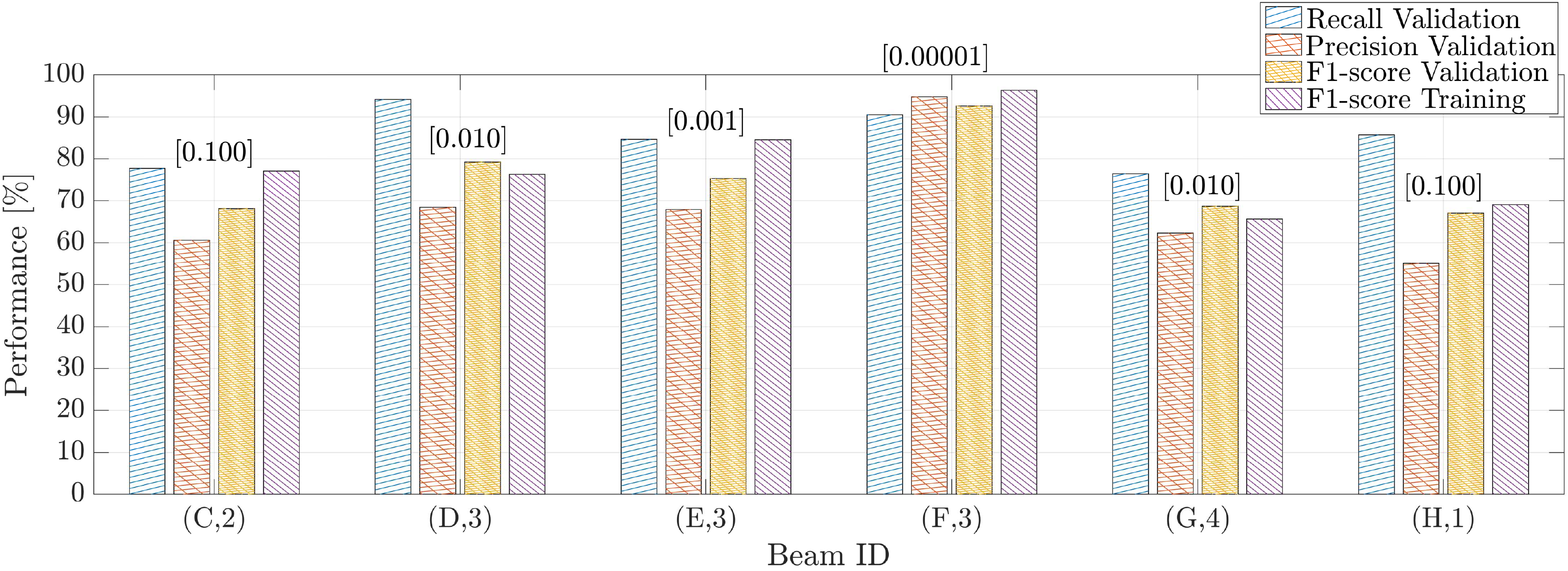}
    	\caption{Training results of the \ac{DNN} models for blockage state prediction for several beams of \ac{BS}-3.}%
    	\label{fig:modelResultsf1score}%
    \end{figure*}%
	
	\emph{2) Dynamic blockage generation:} to model the blockage loss, we adopt the Blockage Model B that is incorporated within the \ac{3GPP} 3D \ac{SCM}. Compared to the stochastic Blockage Model A \cite{8761104}, the deterministic Blockage Model B supports spatial consistency, which enables to simulate the smooth and continuous-time evolution of the blockage effect while capturing the correlated behaviour of the blockage loss for \acp{UE} located closely \cite{3gpp.38.901}. 
	As a case study, we consider the presence of one blocker at a time moving in the network scenario. 
	The blocker has dimensions 2~m $\times$ 3~m and constant speed $v$ selected randomly between two values $1$ m/s and $2$ m/s, in line with the recommended blocker parameters listed in \cite{3gpp.38.901}. 
	The initial position is set at the x-y coordinates $(-20,0)$ m, corresponding to the left side of the network layout, and it moves along the x-axis line from left to right till reaching x-y coordinates $(20,0)$.
	When the blocker arrives at the right side of the network layout, it regenerates at the initial position and repeats the same movement.\footnote{This use case is indicated for scenarios such as future industrial factories, where Automated Guided Vehicles (AGVs) or industrial robots, may maintain similar movement patterns and follow a dedicated pathway in the factory corridor.} 
	
    \emph{3) Beam-specific dataset:} we generate a training dataset per beam, collecting $\snr$ measurements over time for successive simulation steps and repeat the same process for $100$ network drops, including both values of the blocker speed in the same dataset. 
    For instance, the beam $(\rm{C},2)$ dataset has $\mathord{\sim}1.5$ million samples divided between $\mathord{\sim}{80}$ thousand samples for the blocked class and $\mathord{\sim}1.42$ million samples for the non-blocked class. 
    We split this data according to the commonly used ratio of $80$/$20$ to use $80\%$ of the dataset for training the \ac{DNN} model and the remaining $20\%$ of the dataset to validate the \ac{DNN} model performance. 
    During the training, we consider mini-batch of size $N_{bs}=1000$ samples and $50$ training epochs. 
    
    \emph{4) Validation metrics:} as shown before for the beam $(\rm{C},2)$, the number of non-blocked samples significantly outweighs the blocked samples. Thus, for our specific scenario, we validate the \ac{DNN} models adopting the F1-score metric, which is more appropriate than the Accuracy metric for the problems presenting highly unbalanced class-distribution  \cite{5128907}. 
    The F1-score combines through the harmonic mean Precision and Recall and can be calculated as $\text{F1-score}=\frac{2 \times \text{Precision} \times \text{Recall}}{\text{Precision} + \text{Recall}}$. 
    Recall indicates the proportion of actual blocked samples (\ac{TP} cases) over all the blocked samples (\ac{TP} and \ac{FN} cases), and is calculated as $\text{Recall}=\frac{\rm{TP}}{\rm{TP} + \rm{FN}}$. 
    Conversely, Precision indicates the proportion of actual blocked samples among all predicted blocked samples (\ac{TP} and \ac{FP} cases) and is calculated as $\text{Precision} = \frac{\rm{TP}}{\rm{TP} + \rm{FP}}$.  
    
    \emph{5) Prediction parameters:} we fix the prediction window length to $\eta=400$ ms, which is larger than the combined duration of the beam sweeping procedure, i.e. $T_{sweep}=330$ ms, and the handover time, i.e. $T_{HO}=50$ ms \cite{8812724}. 
	We adjusted the \ac{DNN} models input size after several training attempts, which led us to use the number of input beams $L=5$ and set the sliding window duration to $\epsilon=2$ s. 
	
	\subsection{Multi-beam Prediction Results}
    \label{Sec:validationRes}
    In Fig.~\ref{fig:modelResultsf1score}, we show the training and validation results of the \ac{DNN} models taking as an exemplary reference \ac{BS}-3 and several of its blocked beams having different azimuth and elevation \acp{AoD}. 
    Nevertheless, the same process can be repeated without loss of generality for other \acp{BS} of the network. 
    The training process highlighted that the \ac{DNN} models of different beams should be tuned separately. 
    In other words, it is not possible to use the same set of hyper-parameters for all \acp{DNN}. Hence, we perform an independent validation of the \ac{DNN} models beam by beam. 
    During the training, we change the L2 regularisation value in the range $\{10^{-5}-10^{-1}\}$ to reduce the overfitting. 
    We measure the performance of each beam for five different random initialisation of the neuron weights. Then, we save the model that returns the maximum F1-score performance on the validation data. 
    
    Looking at the results in Fig.~\ref{fig:modelResultsf1score}, we report between parenthesis the value of L2 regularisation hyper-parameter that corresponds to the model attaining the best F1-score. 
    Then, we report Recall, Prediction and F1-score results measured on the validation data and the F1-score result measured on the training data. 
    Firstly, most of the beams show Recall validation performance higher than Precision validation performance. 
    This is due to the class weights applied to the loss function, shown in Eq.~\eqref{eq:lossfunction}, which penalises more the \ac{FN} errors than the \ac{FP} errors. 
    This setting increases the probability of predicting the blockage correctly but is likely to make more \ac{FP} errors for the non-blocked samples. 
    Thus the models achieve high Recall while sacrificing their Precision. 
    Secondly, Fig.~\ref{fig:modelResultsf1score} shows that the beam achieves F1-score validation results close to the F1-score training results. 
    This indicates the models' ability to generalise from the training data to other data generated with the simulator assuming different network drops, the same blocker size/trajectory and blocker speed that can vary randomly between the values of $1$ m/s and $2$ m/s, as we will see in Sec.~\ref{Sec:timeSeries} and Sec.~\ref{Sec:throughputRes}, where we deploy the \ac{DNN} models into the simulator. 
	
	\subsection{Evolution of the SNR Time Series}
	\label{Sec:timeSeries}
	
	\begin{figure}[t]
    	\centering
    	\subfloat[$\snr$ and blockage state for blocker speed 1 m/s.]{\includegraphics[width=0.45\textwidth]{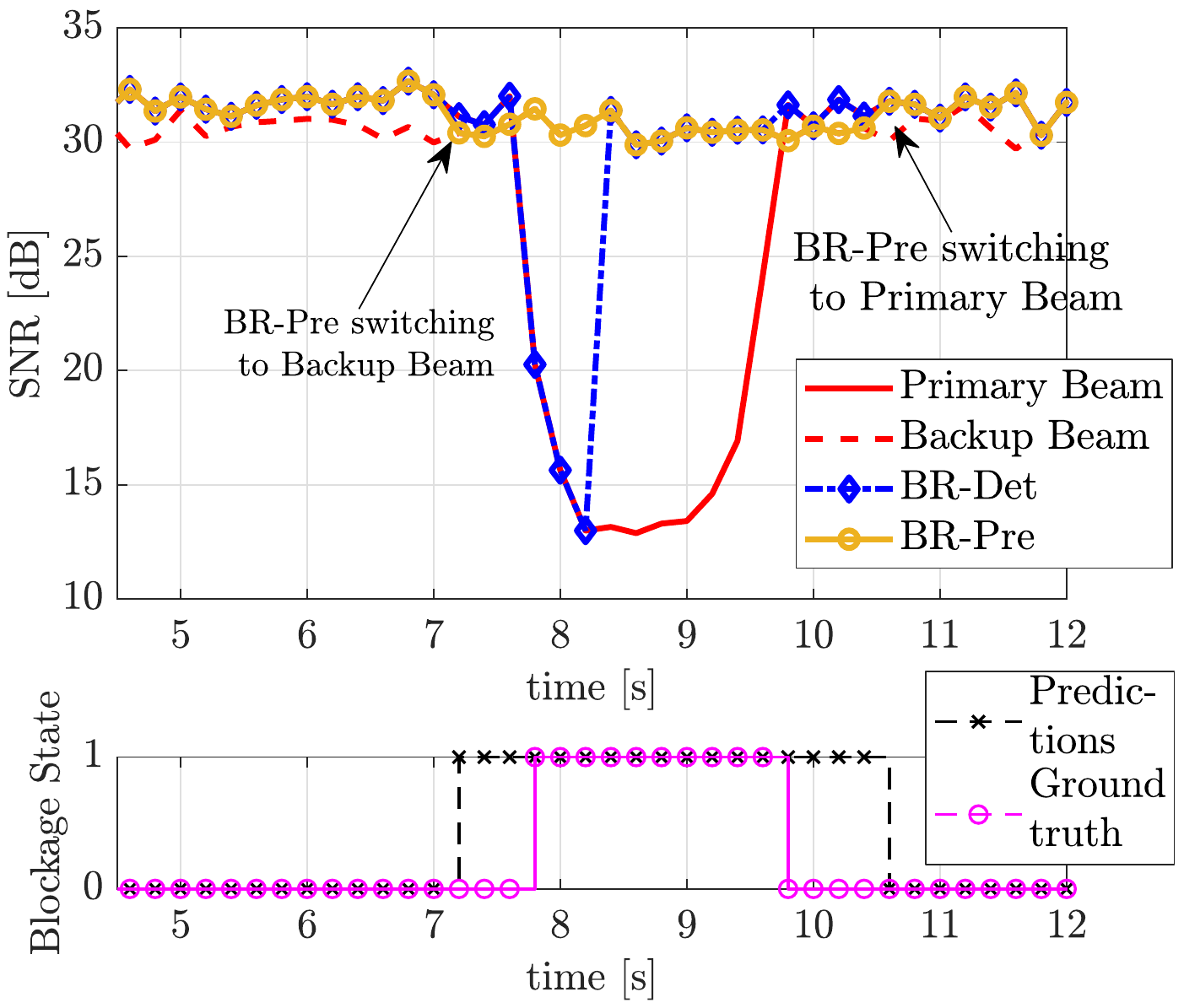}\label{fig:SNRtimeSeries1ms}}%
	    \vspace*{3mm}
    	\\
    	\subfloat[$\snr$ and blockage state for blocker speed 2 m/s.]{\includegraphics[width=0.45\textwidth]{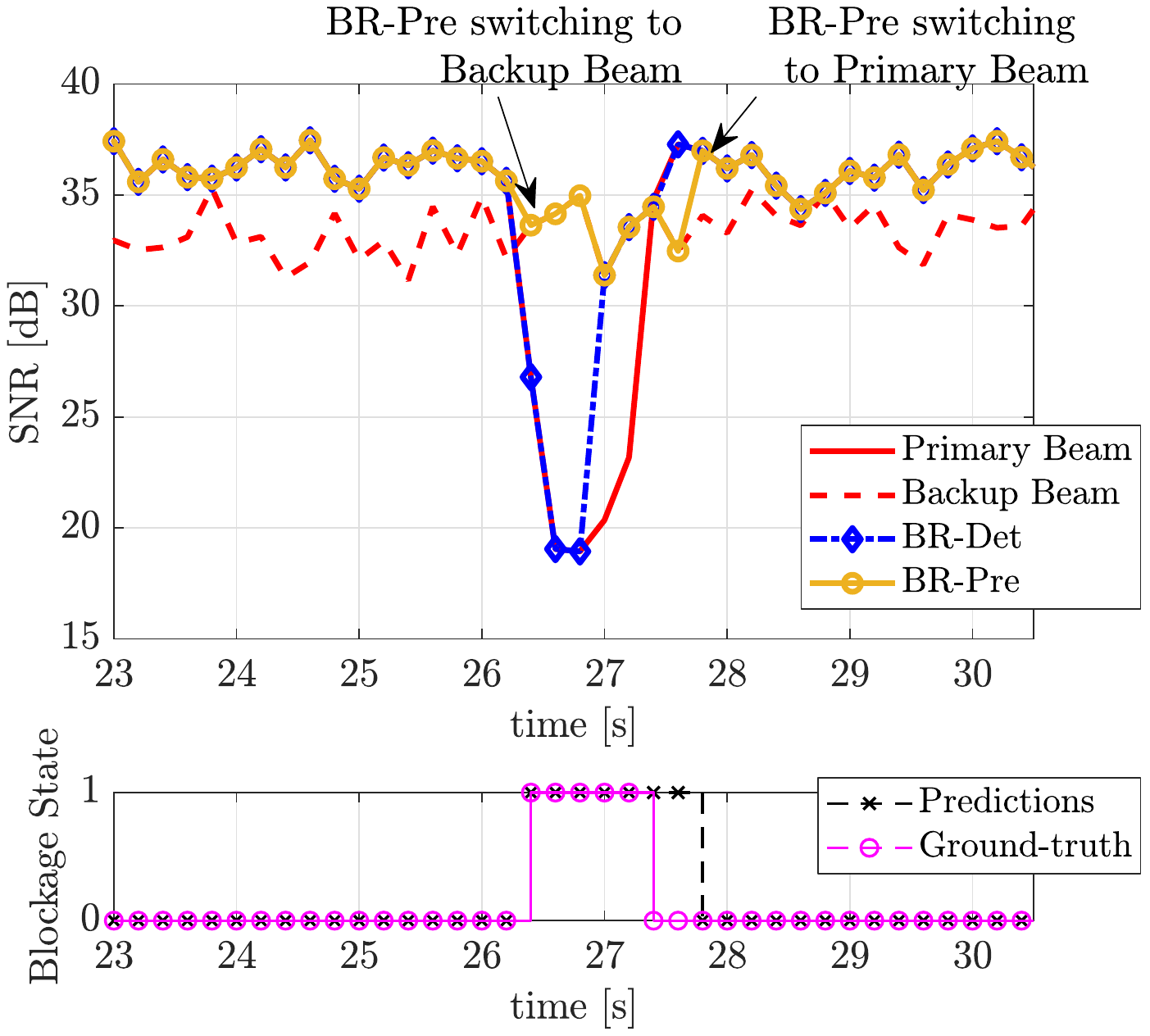}\label{fig:SNRtimeSeries2ms}}%
    	\caption{Comparison between the temporal variations of the $\snr$ for the beam recovery methods (top) and blockage state predictions (bottom) for two different beams and blocker speeds.}
    	\label{fig:timeSeriesSNR}
    \end{figure}

    In this section and in the one that follows, we test the \ac{DNN} models online into the system-level simulator by generating ten different network drops as compared to the ones used in the training dataset with the blocker moving at the speeds $1$ m/s, and other ten network drops with the speed of $2$ m/s. 
    Fig.~\ref{fig:timeSeriesSNR} shows the $\snr$ measurements' evolution in the time domain for two \ac{Tx} beams of the \ac{BS}-3's codebook with the blocker moving at the speeds $1$ m/s and $2$ m/s. 
    These results highlight the BR-Pre behaviour over time and show the differences between the BR-Pre method, the BR-Det method and the case where the transmission continues to be handled by the primary beam. 
    Focusing on Fig.~\ref{fig:SNRtimeSeries1ms}, the bottom part shows the \ac{GT} and predicted primary beam states. 
    It is worth recalling that the \ac{GT} state is determined by considering the intersection between the direct path that joins \ac{Tx} and \ac{Rx} and the rectangular screen modelling the blocker, as described in Section \ref{sec:dataset_generation}. 
    Conversely, the upper part of Fig \ref{fig:SNRtimeSeries1ms} shows the corresponding evolution of the measured $\snr$. 
    
    During the non-blocked time instants, the backup beam pair shows a lower $\snr$ than the primary beam pair because it is established with a secondary \ac{BS}. Differently, before the primary beam enters into a blocked state (as indicated by the ground-truth state in the bottom part of the figure), the $\snr$ of the primary beam pair starts decreasing, while the $\snr$ of the backup beam pair remains stable, providing a higher $\snr$ than the primary beam for the following blocked time instants. 
    The BR-Det method switches to the backup beam with a delay $\beta_1$ that considers the beam sweeping interval initiated after detecting the blockage. Thus, as shown in Fig.~\ref{fig:SNRtimeSeries1ms}, during the time interval $7.8$ s - $8.4$ s, the BR-Det experiences a significant drop of the $\snr$ that is recovered only after switching to the backup beam at the time $t=8.4$ s.
    
    On the other hand, as shown in the two Figs. \ref{fig:SNRtimeSeries1ms} and \ref{fig:SNRtimeSeries2ms}, the BR-Pre method uses the \ac{DNN} models predictions that infer the beam state $\eta$ time instants ahead of the time instant $t$. Thus, the BR-Pre method initiates the beam sweeping procedure at the beginning of the prediction window, i.e. at the time $t-\eta$. 
    After the end of the prediction window, i.e. at the time $t$, when the $\snr$ of the primary beam pair starts decreasing, the BR-Pre method has already switched to the backup beam and avoids the $\snr$ drop that happens with the BR-Det method. 
    However, as shown in Fig.~\ref{fig:SNRtimeSeries1ms} during the time intervals $7.2$ s - $7.6$ s and $9.8$ s - $10.4$ s and in Fig.~\ref{fig:SNRtimeSeries2ms} for the interval $27.4$ s - $27.6$ s the BR-Pre method presents a $\snr$ lower
    than the BR-Det method as the BR-Pre method uses the backup beam instead of the primary beam. 
    This is caused by the \ac{FP} errors associated with the cases when the \ac{DNN} models wrongly predict the beam state as blocked when there is no blockage present. 
    We will show later in Sec. \ref{Sec:throughputOnline} the limited impact of these errors on the data rate performance.
	
	\subsection{Validation of the BR-Pre Method Performance with Online Predictions}\label{Sec:throughputOnline}
	\begin{figure*}[t]
    	\centering
    	\subfloat[CDF of \acp{UE} data rate for blocked time instants.]{\includegraphics[width=0.5\textwidth]{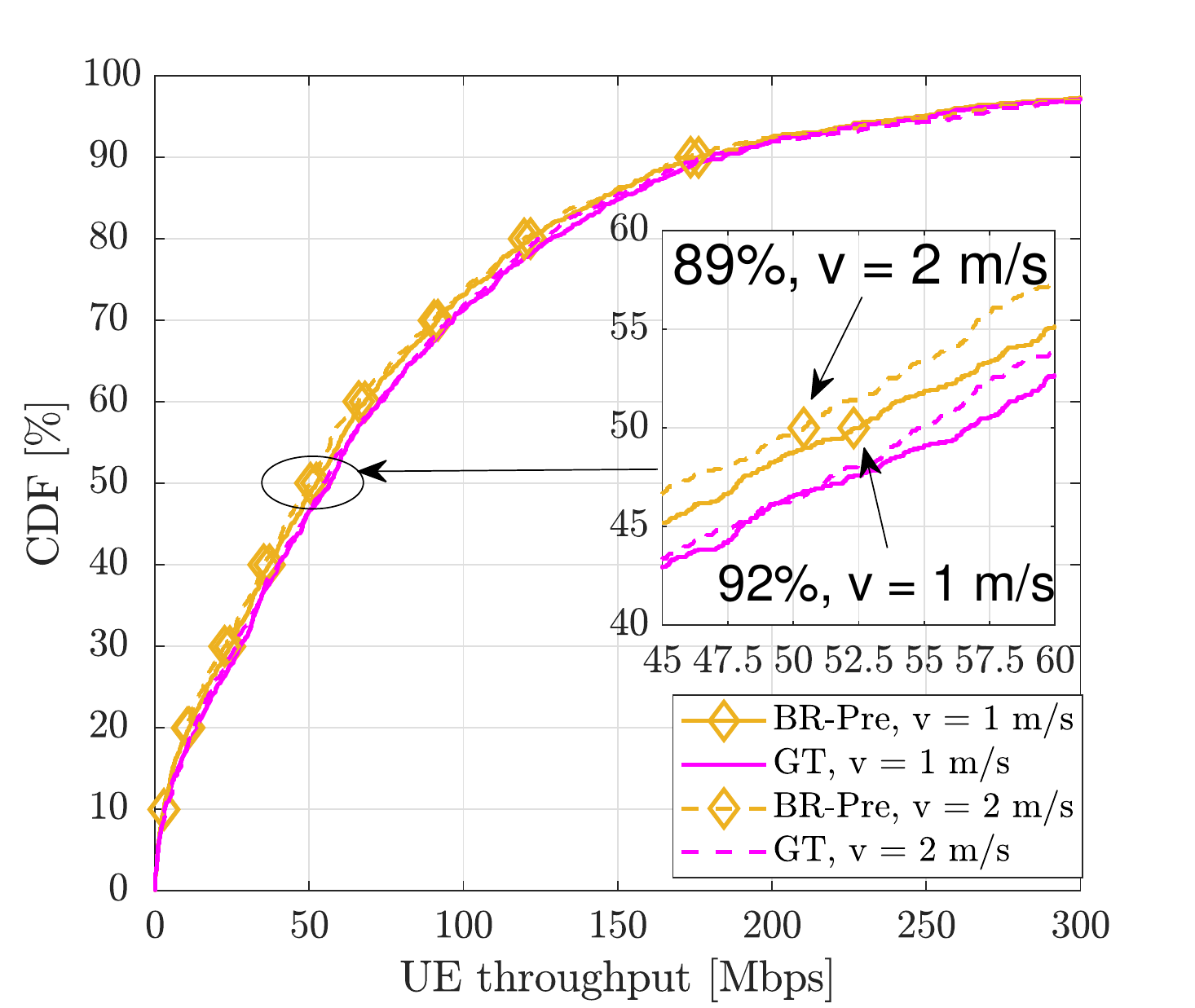}\label{fig:cdfUEthroughputBlockedBRPre_GT}}%
    	\hfil
    	\subfloat[CDF of \acp{UE} data rate for non-blocked time instants.]{\includegraphics[width=0.5\textwidth]{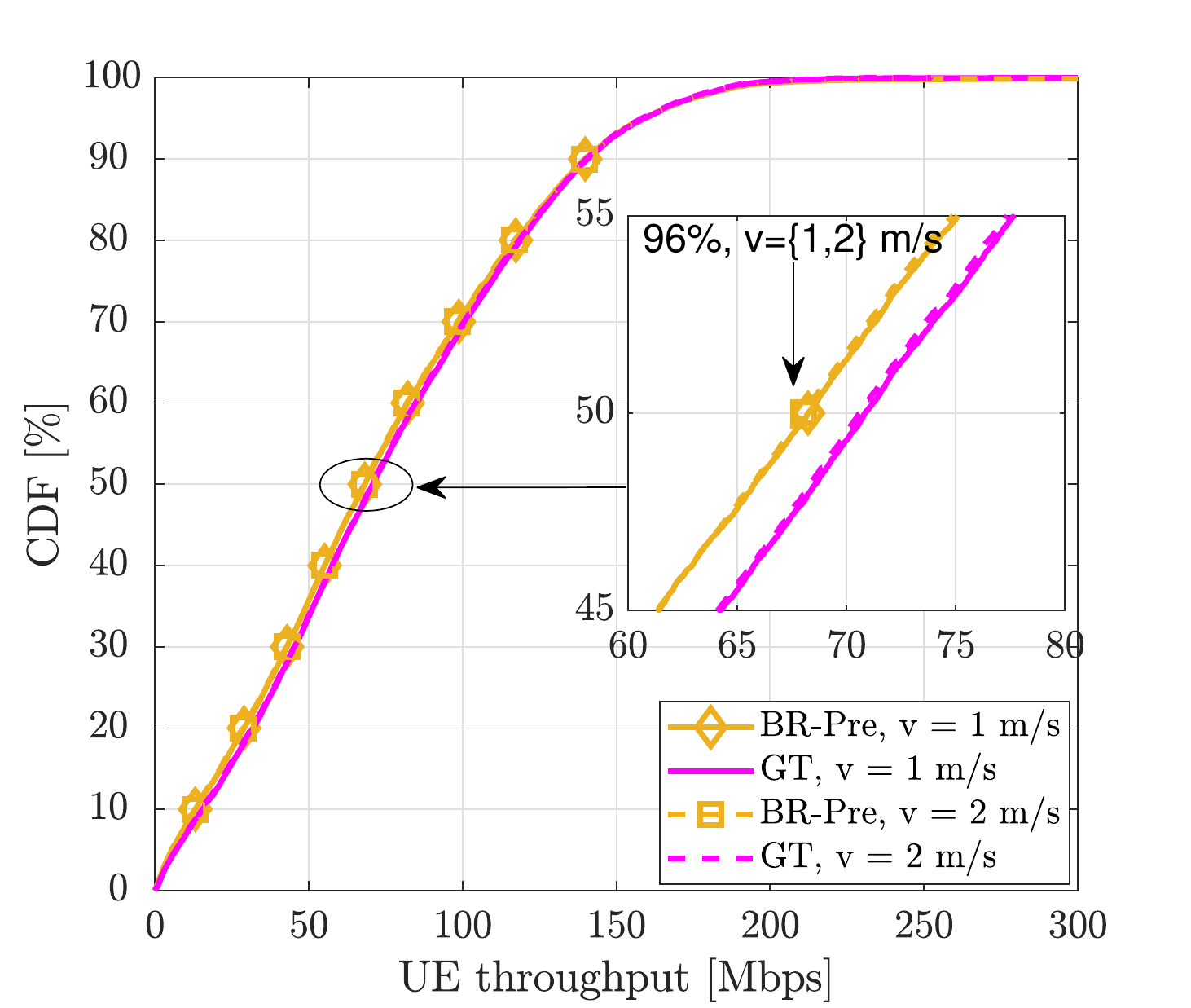}\label{fig:cdfUEthroughputNonBlockedBRPre_GT}}%
    	\caption{Data rate performance of the BR-Pre method with online predictions against \ac{GT} method.}
    	\label{fig:cdfUEthroughputBRPre_GT}
    \end{figure*}
    
    In this section, we show the \acp{CDF} of the \ac{UE} data rate offered by our exemplary \ac{BS}-3 with the blocker moving at speeds $1$ m/s and $2$ m/s. 
    In Fig.~\ref{fig:cdfUEthroughputBRPre_GT}, we show the performance of the BR-Pre method, which uses the \ac{DNN} models to make online predictions, as explained in Sec.~\ref{sec:blockage prediction}, and we compare the results to the \ac{GT} method, representing the upper bound rate achievable when Eq.~\eqref{eq:datarate_withprediction} has perfect knowledge of the future beam states. 
    During the blocked intervals, the \acp{CDF} of Fig.~\ref{fig:cdfUEthroughputBlockedBRPre_GT} show that for the blocker speed of $1$ m/s, the median BR-Pre data rate performance reaches $92\%$ of the \ac{GT} method performance, while for the blocker moving at 2 m/s, the BR-Pre method achieves $89\%$ of the \ac{GT} data rate. The differences between BR-Pre and GT method data rates are due to the \ac{FN} errors that cause the BR-Pre method to use the primary beam during the blocked time instants. Nevertheless, these results show that the BR-Pre method almost matches the GT method data rate performance. 
    
    On the other hand, Fig.~\ref{fig:cdfUEthroughputNonBlockedBRPre_GT} shows the \ac{CDF} of the BR-Pre method data rate achieves $96\%$ of the median GT data rate during non-blocked time instants due to the impact of the \ac{FP} errors. The FP errors cause switching to the backup beam during the non-blocked intervals while the primary beam has a larger \ac{SNR} than the backup beam. 
    Fig.~\ref{fig:cdfUEthroughputNonBlockedBRPre_GT} show that the different \acp{CDF} curves overlap, meaning that the \ac{FP} errors observed in Fig.~\ref{fig:timeSeriesSNR} do not significantly impact the data rate performance of the BR-Pre method. 
	Overall, the results in Figs. \ref{fig:cdfUEthroughputBlockedBRPre_GT} and \ref{fig:cdfUEthroughputNonBlockedBRPre_GT} indicate that the \ac{DNN} models, when deployed online, generalise well for both the blocker speeds. 
	\subsection{Data Rate Performance Comparison Between BR-Pre, BR-Det and BF Methods}\label{Sec:throughputRes}
    \begin{figure*}[t]
    	\centering
    	\subfloat[CDF of \acp{UE} data rate for blocked time instants.]{\includegraphics[width=0.5\textwidth]{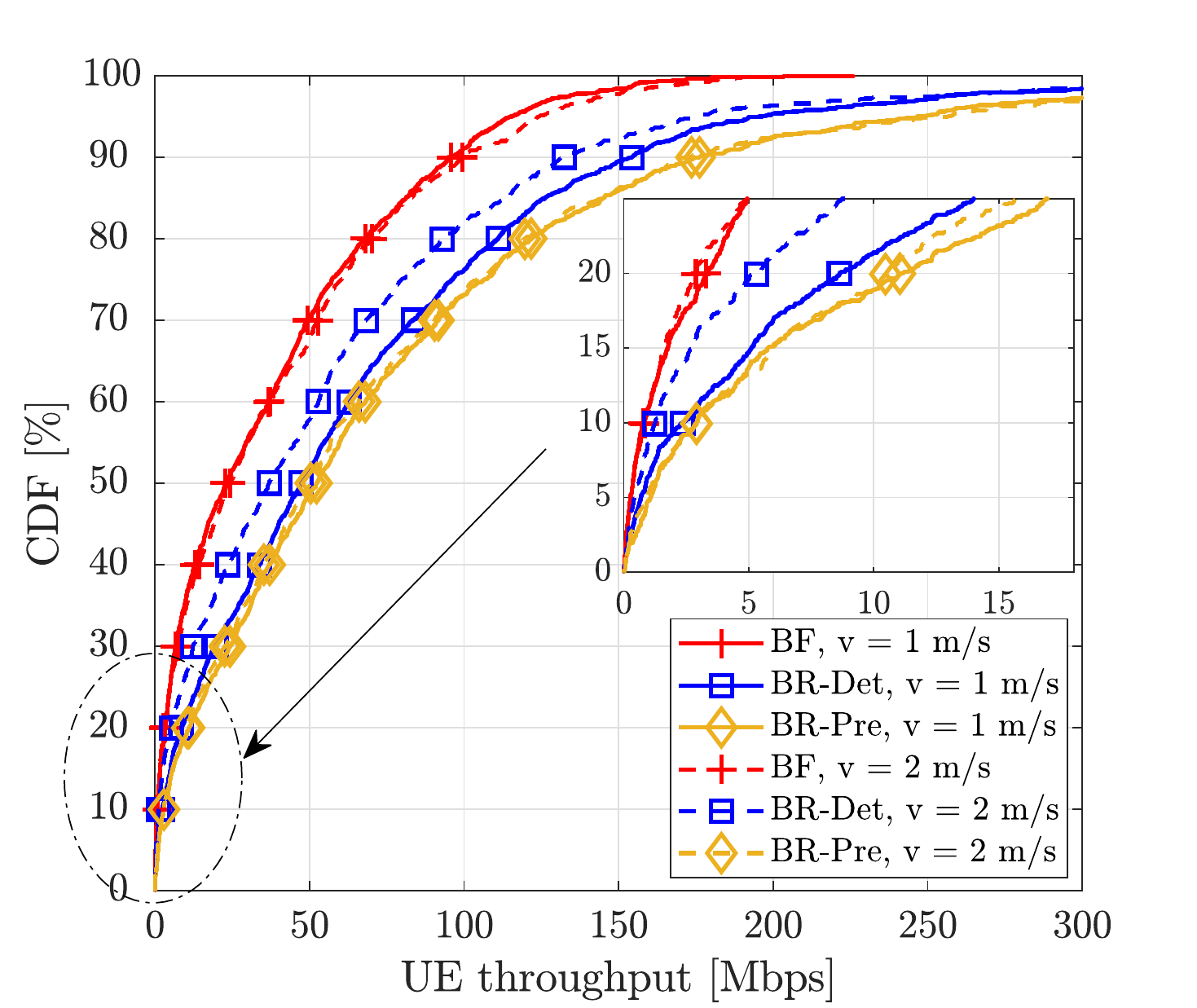}\label{fig:cdfUEthroughputBlockedBF_BRPre_BRDet}}%
    	\hfil
    	\subfloat[CDF of \acp{UE} data rate for non-blocked time instants.]{\includegraphics[width=0.5\textwidth]{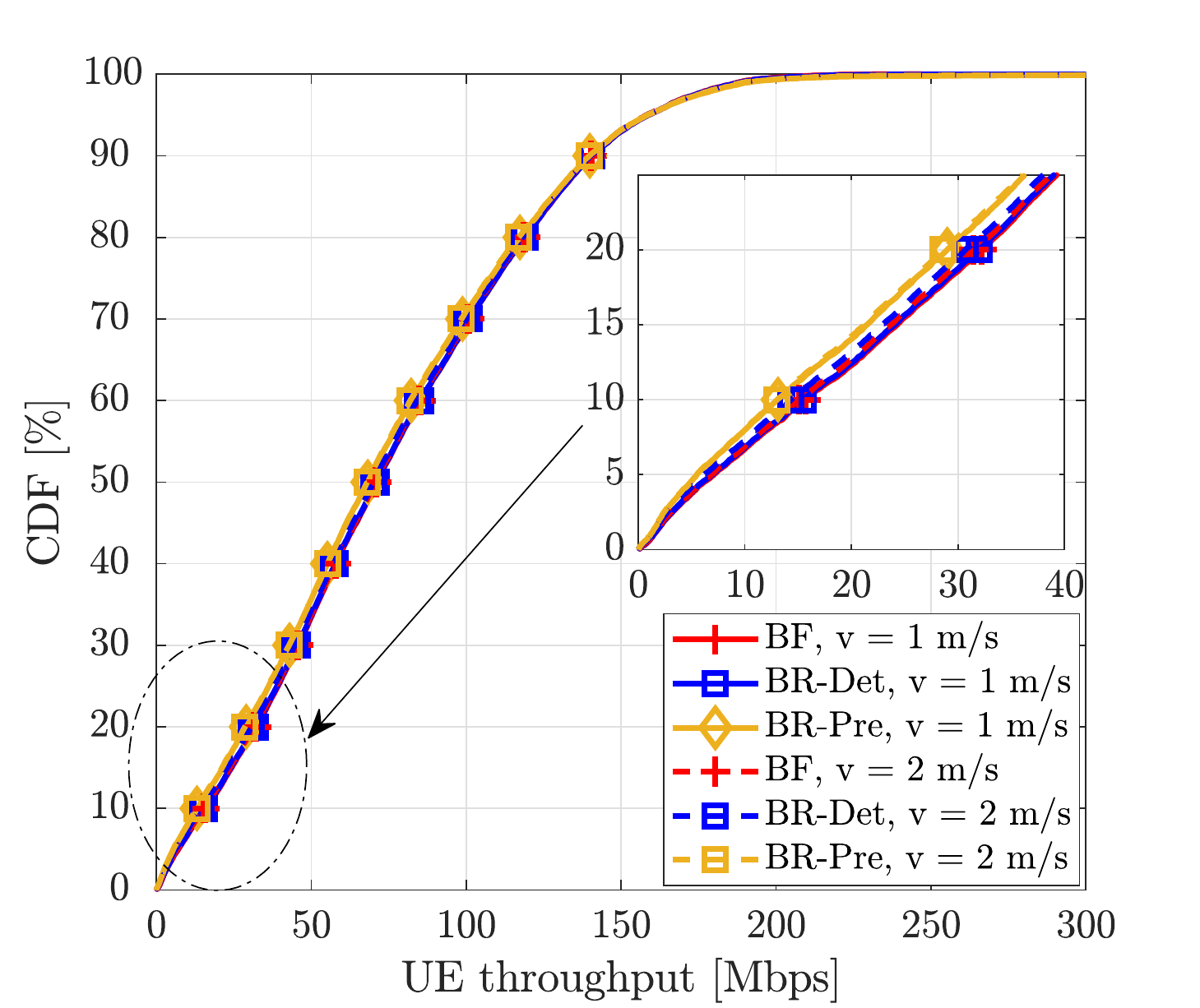}\label{fig:cdfUEthroughputNonBlockedBF_BRPre_BRDet}}%
    	\caption{Data rate performance comparison between BR-Pre, BR-Det and BF methods.}
    	\label{fig:cdfUEthroughputBF_BRPre_BRDet}
    \end{figure*}
    
    In this section, we compare the results between the proposed BR-Pre method, the benchmark BR-Det method reported in Sec.~\ref{sec:blockage detection} and the beam fixed (BF) method, presenting the lower bound rate achievable using the primary beam pair. 
    The \acp{CDF} in Fig.~\ref{fig:cdfUEthroughputBlockedBF_BRPre_BRDet} show that the BR-Pre method outperforms the BR-Det and BF methods during the blocked time instants, especially for the higher blocker speed. 
    These results are explained as follows. We recall that the delay $\beta_1$ for detecting and switching the beam penalises the BR-Det data rate at the start of the blockage event. 
    On the other hand, as shown before in Sec.~\ref{Sec:throughputOnline}, the BR-Pre method performance reaches $92\%$ and $89\%$ of the GT data rate due to the low number of \ac{FN} errors. 
    Overall, the \ac{DNN} models wrong predictions have a minor impact on the data rate performance than the delay introduced for detecting the blockage and switching the beam. 
    At a higher blocker speed, the beam switching happens more often, and the BR-Det data rate is penalised by the delay more times than with the blocker moving at 1 m/s, leading to a more significant difference to the BR-Pre results, as represented in Fig.~\ref{fig:cdfUEthroughputBlockedBF_BRPre_BRDet}. 
    Additionally, Fig.~\ref{fig:cdfUEthroughputNonBlockedBF_BRPre_BRDet} shows that in the non-blocked time instants, the BR-Pre data rates slightly deteriorate compared to the BR-Det data rate by $4\%$ and $3\%$ at the median of the two speeds \acp{CDF}, and by $6\%$ and $5\%$ at the 25-th percentile of the two speeds \acp{CDF}. 
    These results show that during non-blocked time instants, the BR-Pre method performance difference to the BR-Det data rate, caused by the \ac{FP} errors, are marginal. 
    
    \begin{figure*}[t]
    	\centering
    	\subfloat[UE data rate for blocker speed 1 m/s.]{\includegraphics[width=0.44\textwidth]{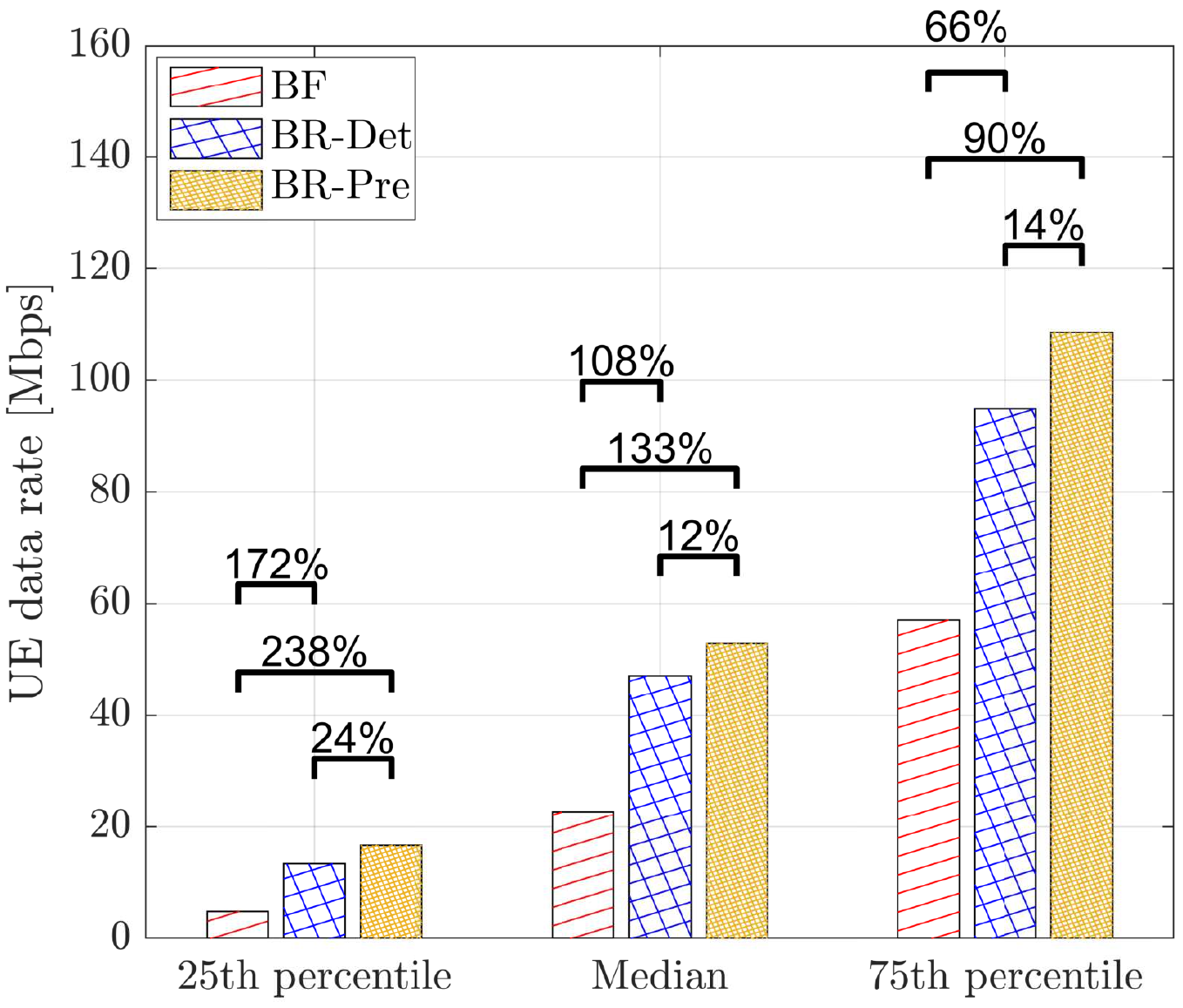}\label{fig:UEthroughput1msBlockedBF_BRPre_BRDet}}%
    	\hfil
    	\subfloat[UE data rate for blocker speed 2 m/s.]{\includegraphics[width=0.44\textwidth]{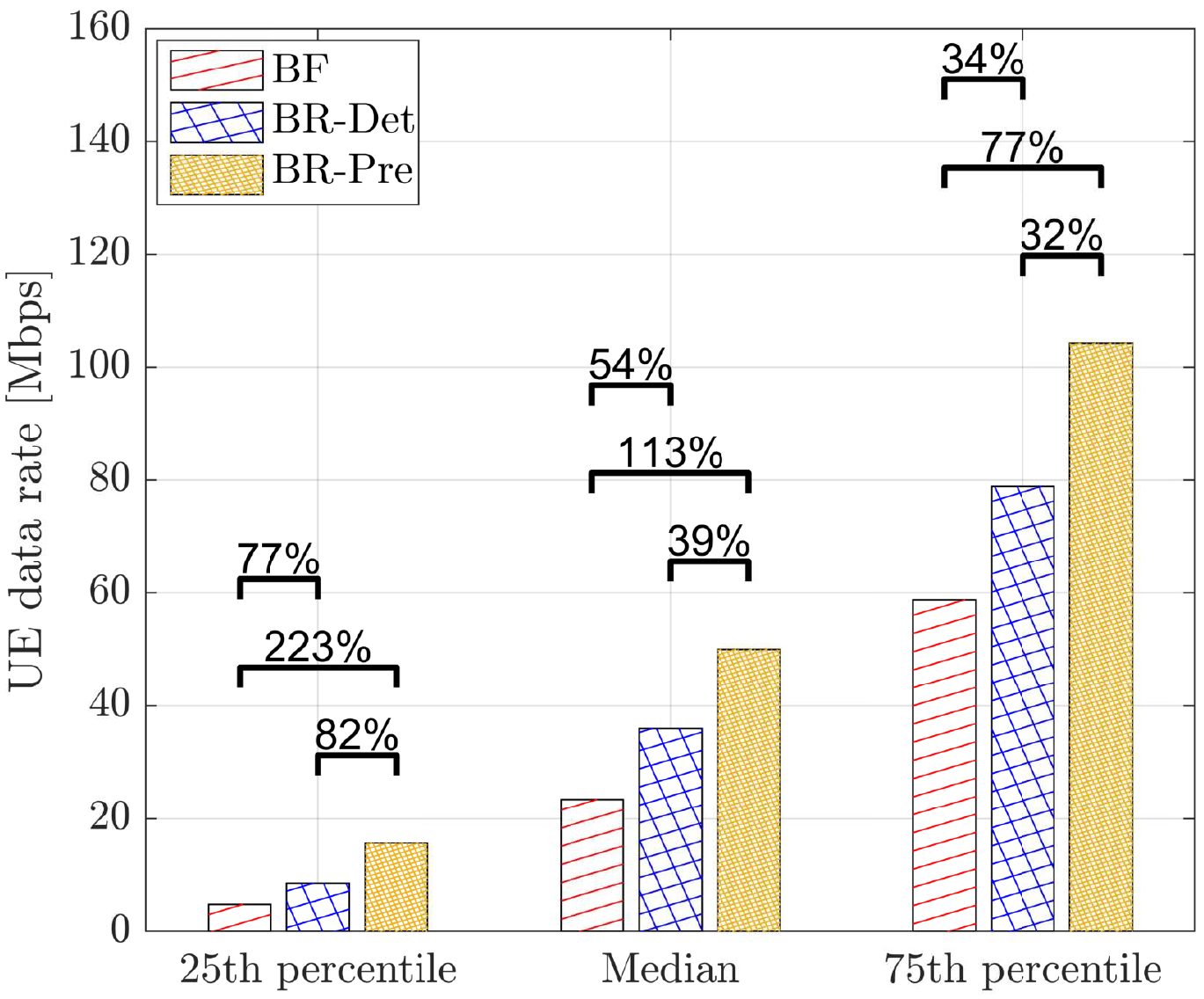}\label{fig:UEthroughput2msBlockedBF_BRPre_BRDet}}%
    	\caption{25-th, 50-th (median) and $75$-th percentiles of the data rate performance for BR-Pre, BR-Det and BF methods during blocked time instants.}\label{fig:UEthroughputBlockedBF_BRPre_BRDet}
    \end{figure*}
    
    In the remainder of this section, we take a deeper look at the data rate performance focusing on the lower part of the \ac{CDF} presented in Fig.~\ref{fig:cdfUEthroughputBF_BRPre_BRDet}, as the worst served \acp{UE} are most likely to suffer in blocked time instants to meet the requirement of the most demanding applications. 
    Focusing on the 25th percentile results at blocker speed $1$ m/s shown in Fig.~\ref{fig:UEthroughput1msBlockedBF_BRPre_BRDet}, the BR-Det and BR-Pre methods improve the BF data rate by $172\%$ and $238\%$ respectively and with BR-Pre outperforming BR-Det by $24\%$.
    Similarly, looking at 25th percentile results of Fig.~\ref{fig:UEthroughput2msBlockedBF_BRPre_BRDet} with the blocker moving at the speed of $2$ m/s, the BR-Det and the BR-Pre methods improve the BF method data rate by $77\%$ and $223\%$, respectively. The advantage of BR-Pre over BR-Det increases, and it is assessed at $84\%$.
    Note that the advantage offered by the BR-Pre method over the BR-Det one is more evident at the higher speed, as the reaction time of BR-Det does not scale with the higher frequency of the blockage events and their shorter duration, thus, by the time the method triggers the switch to the backup beam pair, the drop in the signal level caused by the blockage is either about to finish or already finished. 
    In contrast, the BR-Pre data rate performance does not depend on the blocker speed, showing how predicting in advance the blockage occurrence is effective in indoor factory network deployments. 
    
    Similar considerations apply for the median and the 75th percentiles of \acp{CDF} with blocker speed of $2$ m/s showing $39\%$ and $32\%$ performance gains of our proposed prediction method over the detection one. 
	
	\section{Conclusion}\label{Sec:6}
	
	In this paper, we re-use existing beam measurement report messages from multiple \acp{UE} as input data for beam-specific \ac{DNN} models that predict blockage events for multiple beams of the one exemplary \ac{BS}. 
	Our proposed multi-\ac{UE} prediction-based method utilises the beam state predictions to control and trigger the beam recovery procedure ahead of the blockage events. This enables switching to a backup beam pair before the blockage disrupts the primary beam path. Thus, the prediction-based method allows for more stable signal quality and is more effective in significantly reducing the typical loss of data rate during the blockage event shown by commonly used methods based on detection. 
    The validation of the proposed method with online predictions shows how data rate performance achieves $92\%$ and $89\%$ of the GT data rate for the blockers speeds of $1$ m/s and $2$ m/s, respectively, confirming that the results are very close to the ideal case when the system has perfect knowledge of the future beam states.
    Moreover, the low number of \ac{FP} errors rarely triggers the switching to the backup beam during non-blocked intervals and impact the data rate performance marginally, losing only $4\%$ and $3\%$ to the BR-Det method median data rate.
    In addition, we show that the higher blocker speed penalises BR-Det method as it is less effective in reacting to more frequent and shorter blockage events. In contrast, the BR-Pre data rate performance remains close to the GT. 
    Finally, we found that for the worst served \ac{UE} (at 25-th percentile of the \ac{CDF}), the BR-Det improves the BF data rates by $172\%$ and $77\%$ for speeds $1$ m/s and $2$ m/s, while BR-Pre improves BF method data rates by $238\%$ and $223\%$ for speeds $1$ m/s and $2$ m/s. 
    Overall, for the worst 25-th percentile \ac{UE}, the proposed BR-Pre method improves the the BR-Det method data rates by $24\%$ at a blocker speed of $1$ m/s and $82\%$ at $2$ m/s.  
    Future works will focus on extending the current blockage scenario to diversify the blockage movement's trajectory and further reducing the time associated to the beam-specific training process. 
    
    \appendices
    
    \section{3D Spatial Channel Model}
    \label{appendix:channelModel}
    We generate the \ac{mmWave} channel following the 3D \ac{SCM} specified in 
    \cite{3gpp.38.901}, which considers a geometry-based stochastic channel model that accounts for a scattering environment formed by $N_C$ clusters individually composed by $N_P$ sub-paths. 
    The channel impulse response between the \ac{Tx}-\ac{Rx} antenna pair of the \ac{BS}-\ac{UE} link can be expressed as \cite{3gpp.38.901,8770245,6824736}
    \begin{equation}\label{eq:channel}
    \begin{split}
    H_{u,s} (t,\tau)
    =\rho\sum_{\ell=1}^{N_C}\sum_{p=1}^{N_P} & \mathbf{F}_{Rx} (\theta_{\ell,p}^A, \phi_{\ell,p}^A)
    e^{j\Phi_{\ell,p}} \mathbf{F}_{Tx}(\theta_{\ell,p}^D, \phi_{\ell,p}^D)\\
    &\times e^{j \left( {\mathbf{k}}_{Rx,\ell,p}^{\sfT}~\cdot~ {\mathbf{d}}_{Rx,u} + {\mathbf{k}}_{Tx,\ell,p}^{\sfT}~\cdot~ {\mathbf{d}}_{Tx,s} \right)}\\
    &\times \sqrt{P_{\ell,p}} \cdot 10^{-\frac{BL_{\ell,p}(t)}{20}}\delta (\tau - \tau_{\ell,p}),
    \end{split}
    \end{equation}
    where $\rho=\sqrt{10^{-\frac{PL+\sigma_{SF}}{10}}}$ represents the slow channel gain, which includes pathloss $PL$ and shadowing $\sigma_{SF}$. 
    For each sub-path $p$ in cluster $\ell$, the model specifies \ac{AoA} ($\theta_{\ell,p}^A, \phi_{\ell,p}^A$) and \ac{AoD} ($\theta_{\ell,p}^D, \phi_{\ell,p}^D$), which modify the \ac{Rx} and \ac{Tx} antenna field patterns $\mathbf{F}_{Rx}$ and $\mathbf{F}_{Tx}$, and $\Phi_{\ell,p}$, which represents a random initial phase if polarisation is not considered. 
    Moreover, the terms $\exp(j{\mathbf{k}}_{Rx,\ell,p}^{\sfT} \cdot {\mathbf{d}}_{Rx,u})$ and $\exp(j{\mathbf{k}}_{Tx,\ell,p}^{\sfT} \cdot {\mathbf{d}}_{Tx,s})$ represent the array responses of the \ac{Rx} and \ac{Tx} antennas, where $\mathbf{k}_{Rx,\ell,p}$ and $\mathbf{k}_{Tx,\ell,p}$ are the \ac{Rx} and \ac{Tx} wave vectors along the directions of the $p$-th sub-path in cluster $\ell$ such that $\norm{{\mathbf{k}}}=\frac{2\pi}{\lambda_0}$.
    Additionally, ${\mathbf{d}}_{Rx,u}$ is the location vector of the receiving antenna $s$ whereas ${\mathbf{d}}_{Tx,s}$ is that of the transmitting antenna $u$ computed in the global Cartesian coordinate system. 
    Finally, for each sub-path $p$ in cluster $\ell$, the model accounts for the power gain $P_{\ell,p}$, blockage loss $BL_{\ell,p}(t)$ and propagation delay $\tau_{\ell,p}$. 
    To note that Eq.~\eqref{eq:channel} represents the channel impulse response for the \ac{NLoS} case.
    Thus, we add the \ac{LoS} channel coefficient to Eq.~\eqref{eq:channel} and we scale both terms by the Rician K-factor $K_{R}$. The resulting channel impulse response for the \ac{LoS} case can be expressed as 
    \begin{equation}
    \begin{split}
    H_{u,s}^{\rm{LoS}} (t,\tau)
    &=\sqrt{\frac{1}{K_{R}+1}} H_{u,s}^{\rm{NLoS}} (t,\tau)\\
    &+\sqrt{\frac{K_{R}}{K_{R}+1}} H_{u,s,1}^{\rm{LoS}} (t)\delta(\tau - \tau_1),
    \end{split}
    \end{equation}
    where $H_{u,s,1}^{\rm{LoS}}(t)$ and $\tau_1$ are the channel impulse response and the propagation delay of the \ac{LoS} path, respectively \cite{8770245}. 
    
    \begin{table*}[t]
    	\centering
    	\caption{3GPP-based system-level simulation parameters}
    	\begin{tabulary}{0.8\columnwidth}{|p{2cm}|p{4cm}|p{8cm}|}
    		\hline
    		\multirow{3}{*}{NR Numerology} & Carrier frequency & 28 GHz \\ \cline{2-3} 
    		& System bandwidth~/~Total \acp{RB} & 396 MHz~/~275 \acp{RB} \cite{3gpp.38.300} \\ \cline{2-3} 
    		& Sub-carrier spacing~/~\ac{TTI} duration & 120 KHz~/~0.125 ms \cite{3gpp.38.300} \\ \hline
    		\multirow{6}{*}{BS description} & Network layout & Room size of 40m $\times$ 50m $\times$ 3m, 4 sites, 3~sectors/site \\ \cline{2-3} 
    		& Deployment & Grid-based with \ac{ISD}: $20$~m, height: $3$~m \\ \cline{2-3} 
    		& Antenna array & \ac{UPA} with element spacing $0.5\lambda$, Number of antennas per array: 8~$\times$~8~=~64, mechanical downtilt: $\ang{20}$, antenna boresigh $\{\ang{30}, \ang{150}, \ang{270}\}$ \cite{3gpp.RT-170019} \\ \cline{2-3} 
    		& Single antenna element pattern & $\ang{90}$ H $\times$ $\ang{90}$ V beamwidths, 5 $\mathrm{dBi}$ max. \cite{3gpp.RT-170019} \\ \cline{2-3} 
    		& Beamforming & Fully-analog architecture based on codebook with size $N_{\mathrm{CB},Tx}=64$ \\ \cline{2-3} 
    		& \ac{Tx} power & 20 dBm \\ \hline
    		\multirow{5}{*}{UE description} & Deployment & Random, 20 \acp{UE}/sector on average, all \acp{UE} served, height: $1$~m \\ \cline{2-3} 
    		& Antenna array & \ac{UPA} with element spacing $0.5\lambda$, Number of antennas per array: 4~$\times$~4~=~16, antenna boresigh $\sim{\mathcal {U}}[\ang{0},\ang{360}]$ \cite{3gpp.RT-170019} \\ \cline{2-3} 
    		& Single antenna element pattern & $\ang{90}$ H $\times$ $\ang{90}$ V beamwidths, 5 $\mathrm{dBi}$ max. \cite{3gpp.RT-170019} \\ \cline{2-3} 
    		& Beamforming & Fully-analog architecture based on codebook with size $N_{\mathrm{CB},Rx}=16$ \\ \cline{2-3} 
    		& Noise figure & 10 dB \\ \hline
    		\multirow{3}{*}{\begin{tabular}[c]{@{}c@{}}Blocker \\ description\end{tabular}} & Model & Geometric-based model B \cite{3gpp.38.901} \\ \cline{2-3} 
    		& Dimensions & 2~m $\times$ 3~m \\ \cline{2-3} 
    		& Trajectory & Linear with speed $v=\{1, 2\}$ m/s \\ \hline
    		\multirow{4}{*}{\begin{tabular}[c]{@{}c@{}}mmWave\\ Channel \\ Description\end{tabular}} & Path loss and LOS probability & 3GPP 3D InH-Open office \cite{3gpp.38.901} \\ \cline{2-3} 
    		& Shadowing & Log-normal with $\sigma$=3 / 8 dB (\ac{LoS} / \ac{NLoS}) \cite{3gpp.38.901} \\ \cline{2-3} 
    		& Fast fading & Ricean with log-normal K-factor \cite{3gpp.38.901} \\ \cline{2-3} 
    		& Thermal noise & -174 dBm/Hz power spectral density \\ \hline
    	\end{tabulary}
    	\label{table:system_level_parameters} 
    \end{table*}
    
    \section{Blockage Model} 
    \label{appendix:blockModel}
    To capture the blockage loss given by blockers moving in the environment, we consider the \ac{3GPP} Blockage Model B, which introduces a time-dependent component $BL(t)$ that reduces the power of the 3D channel cluster's sub-path \cite{3gpp.38.901}. 
    The Blockage Model B deploys a 3D rectangular screen with sizes $w \times h$ for each blocker moving in the scenario and computes the total power loss given by the knife-edge diffraction from the edges of the screen \cite{4653341}. 
    The model maintains consistent results over time, space, and channel frequency components because of its geometric approach. 
    The overall blockage attenuation can be expressed as \cite{3gpp.38.901}
    \begin{equation}\label{blockageLoss}
    BL = -20\log_{10}\left(1-(F_{h_1}+F_{h_2})(F_{w_1}+F_{w_2})\right),
    \end{equation}
    where $F_{h_1}, F_{h_2}, F_{w_1}, F_{w_2}$ represent the diffraction losses observed at the \ac{Rx} corresponding to the four edges of the screen and can be expressed as \cite{3gpp.38.901}
    \begin{equation*}\label{eq:Fw1}
    F_{w_{1,2}|h_{1,2}}\!=\!
    \begin{dcases}
    \frac{\tan^{-1}\!\left(\pm\frac{\pi}{2}\sqrt{\frac{\pi}{\lambda_0}\left(D1_{w_{1,2}|h_{1,2}}\!+\!D2_{w_{1,2}|h_{1,2}}\!-\!d\right)}\right)}{\pi}\\
    \qquad\text{for \ac{LoS} path,}\\
    \frac{\tan^{-1}\!\left(\pm\frac{\pi}{2}\sqrt{\frac{\pi}{\lambda_0}\left(D1_{w_{1,2}|h_{1,2}}-d'\right)}\right)}{\pi}\\
    \qquad\text{for \ac{NLoS} path,}
    \end{dcases}
    \end{equation*}
    where $\lambda_0$ is the wavelength, $d$ and $d'$ are the \ac{BS}-\ac{UE} and blocker-\ac{UE} distances, corresponding to the LoS and NLoS paths, respectively. 
    $D1_{w_{1,2}|h_{1,2}}$ and $D2_{w_{1,2}|h_{1,2}}$ denote the line segments connecting the screen edges to the \ac{Tx} or \ac{Rx} points. 
    These distances are evaluated in the top view for $w1$ and $w2$ and in the side view for $h1$ and $h2$. 
    If the screen intersects the \ac{Rx} path, the $+$ sign is applied at both edges. Differently, if the screen does not intersect the \ac{Rx} path and one of the edges is still diffracting the signal, the $-$ sign is applied to the edge closest to the \ac{Rx} path and the $+$ sign is applied to the edge farthest from the \ac{Rx} path. 
    
    \section{3GPP-based system-level setup}\label{appendix:simulationParameters}
    Table~\ref{table:system_level_parameters} provides the set of simulation parameters used to configure the system-level simulator. 

	\bibliographystyle{IEEEtran}
	\bibliography{mycollection}
\end{document}